# Reconstructing antibody repertoires from error-prone immunosequencing datasets

Alexander Shlemov[1,*], Sergey Bankevich[1,*], Andrey Bzikadze[1,2], Maria A. Turchaninova[3], Yana Safonova[1,**] and Pavel A. Pevzner[1,4]

[1]Center for Algorithmic Biotechnology, Institute for Translational Biomedicine, St. Petersburg University, St. Petersburg, Russia

[2]Department of Statistical Modelling, St. Petersburg University, St. Petersburg, Russia

[3]Institute of Bioorganic Chemistry, Russian Academy of Sciences, Moscow, Russia

[4]Department of Computer Science and Engineering, University of California, San Diego, USA

*These authors contributed equally to this work
**corresponding author, safonova.yana@gmail.com

## Abstract

Transforming error-prone immunosequencing datasets into antibody repertoires is a fundamental problem in immunogenomics, and a prerequisite for studies of immune responses. Although various repertoire reconstruction algorithms were released in the last three years, it remains unclear how to benchmark them and how to assess the accuracy of the reconstructed repertoires. We describe a novel IGREC algorithm for constructing antibody repertoires from high-throughput immunosequencing datasets and a new framework for assessing the quality of reconstructed repertoires. Benchmarking IGREC against the existing antibody repertoire reconstruction tools has demonstrated that it results in highly accurate repertoire reconstructions. Surprisingly, antibody repertoires constructed by IGREC from barcoded immunosequencing datasets in blind mode (without using unique molecular identifiers information) improved upon the repertoires constructed by the state-of-the-art tools that use barcoding. This finding suggests that IGREC may alleviate the need to generate repertoires using the barcoding technology (the workhorse of current immunogenomics efforts) because our computational approach to error correction of immunosequencing data ends up being nearly as powerful as the experimental approach based on barcoding.

## Introduction

Recent progress in sequencing technologies enabled high-throughput generation of full-length antibody sequences and opened a possibility to assess the diversity of the antibody repertoires in various species (Georgiou et al. 2014; Robinson, 2015; Yaari and Kleinstein 2015; Greiff et al. 2015). However, transforming an error-prone immunosequencing dataset (*Rep-seq* dataset) into an accurate antibody repertoire is a challenging problem (Vander Heiden et al. 2014; Bolotin et al. 2015; Safonova et al. 2015) that is a prerequisite for a multitude of downstream studies of adaptive immune systems (Figure 1) which include reconstruction of clonal lineages (Gupta et al. 2015;

Briney et al. 2016), analysis of immune response dynamics (Galson et al. 2016; Liu et al. 2016; Laserson et al. 2014), analysis of recombination events and secondary diversification (Murugan et al. 2012; Elhanati et al. 2015), immunoproteogenomics (Kindi et al. 2016; Lavinder et al. 2015; Safonova et al. 2015; Cheung et al. 2012), and population analysis of Ig germline genes (Gadala-Maria et al. 2015).

The standard Rep-seq protocols include an amplification step that introduces *amplification errors* that can be propagated by consecutive amplification cycles, thus generating pseudo-diversity of an antibody repertoire (Pienaar et al. 2006; Bolotin et al. 2012). These errors are further compounded by *sequencing errors* in reads (estimated at 0.5% per base on average in Illumina reads), leading to error-prone Rep-seq datasets, which trigger errors in the constructed repertoires, and complicate downstream analysis of antibodies (Figure A1). Thus, *error correction* of Rep-seq datasets with follow-up repertoire construction is a prerequisite for any downstream analysis of immunosequencing data.

The *Unique Molecular Identifiers (UMIs)* barcoding technology (Kinde et al. 2011; Kivioja et al. 2011) allows one to analyze low-abundant receptor sequences and to error-correct amplification errors (Vollmers et al. 2013; Shugay et al. 2014; Cole et al. 2016; Turchaninova et al. 2016; de Bourcy et al. 2017). Combinations of molecular and *cell barcoding*, such as *linkage RT-PCR* (DeKosky et al. 2013; DeKosky et al. 2014; Robinson 2015) or *droplet barcodes* (McDaniel et al. 2016), allow one to analyze chain pairing in antibodies and TCRs. These immunosequencing protocols have different requirements with respect to repertoire construction algorithms; e.g., while aggressive error-correction works well for barcoded Rep-seq datasets, it results in overcorrection and loss of natural diversity for non-barcoded datasets.

Construction of full-length antibody repertoires is a more difficult computational problem than the well-studied *TCR repertoire construction* problem (Kuchenbecker et al. 2015; Gerritsen et al. 2016), the *V(D)J classification* problem (Ye et al. 2013; Gaëta et al. 2007; Elhanati et al. 2015), and the *CDR3 classification* problem (Robins et al. 2009; Freeman et al. 2009; Robins et al. 2010; Warren et al. 2011). In fact, CDR3 classification and full-length antibody repertoire construction are two different clustering problems with increasing granularity of partition into clusters and with different biological applications. We note that the CDR3 classification can be derived from a full-length antibody repertoire, but not *vice versa*.

Recently, several tools for constructing full-length antibody repertoires have been developed, including MIGEC (Shugay et al. 2014), PRESTO (Vander Heiden et al. 2014), MIXCR (Bolotin et al. 2015), and IGREPERTOIRECONSTRUCTOR (Safonova et al. 2015). However, all these tools have limitations making it difficult to benchmark them. Moreover, assessing the quality of the constructed antibody repertoires and benchmarking various repertoire construction algorithms remain a challenge since it is unclear how to construct the *reference antibody repertoire*.

Benchmarking various computational tools in genomics would not be possible without the development of specialized quality assessment tools aimed at various applications. For example, benchmarking of assembly tools would not be possible without the development of such tools for evaluating genomics, metagenomics, and transcriptomics assemblies as QUAST by Gurevich et al. (2013), METAQUAST by Mikheenko et al. (2016), and RNAQUAST by Bushmanova et al. (2016) respectively. We believe that, similar to other areas of genomics, developing a benchmarking framework for repertoire reconstructions is a pre-requisite for objective evaluation of the state-of-art immunoinformatics algorithms. However, assessing the quality of antibody repertoires remains a poorly addressed challenge, making it nearly impossible to compare various repertoire construction tools.

We present the IGREC tool for antibody repertoire construction from both barcoded and non-barcoded immunosequencing datasets and the IGQUAST tool for quality assessment of antibody repertoires. We demonstrate that IGREC results in accurate repertoire reconstructions on diverse Rep-seq datasets. Moreover, repertoires constructed by IGREC from barcoded Rep-seq datasets in blind mode (without using barcoding information) improved on the repertoires constructed by the state-of-the-art tools that use barcoding information. This surprising finding suggests that advanced repertoire construction algorithms may alleviate the need to generate barcoded repertoires, thus reducing experimental effort and potentially changing the currently dominant barcoding-based immunosequencing protocols.

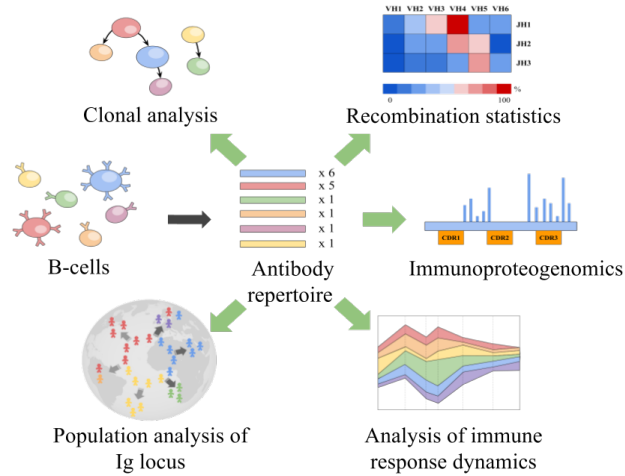

**Figure 1: Antibody repertoire reconstruction is a prerequisite for solving various immunogenomics problems.** Accurate repertoires are needed for (i) reconstruction of clonal lineages, (ii) analysis of immune response dynamics, (iii) analysis of recombination events and secondary diversification, (iv) immunoproteogenomics, and (v) population analysis of the Ig germline genes.

# Methods

**IGREC pipeline**

The IGREC tool addresses the deficiencies in IGREPERTOIRECONSTRUCTOR by Safonova et al. (2015), which is limited to non-barcoded data and is very time- and memory-consuming. IGREC generates an antibody repertoire by partitioning error-prone immunosequencing reads (covering the entire variable regions of immunoglobulins) into clusters. The goal is to place reads from the same antibody into the same cluster, while placing reads from different antibodies into different clusters. This results in a difficult clustering problem since reads from the same antibody differ by sequencing and amplification errors, and the number of these errors often exceeds the number of differences between antibodies from different clusters. We define the *antibody sequence* as the consensus of reads in a cluster, and its *abundance* as the number of reads in a cluster.

Since Rep-seq datasets often contain various transcripts from non-immunoglobulin genes, IGREC first aligns all reads against the database of immunoglobulin *germline genes* and filters out the unaligned reads. Alignment of immunosequencing reads against the germline database or *V(D)J labeling* is a well-studied problem (Ye et al. 2013; Gaëta et al. 2007). Instead of performing full-scale V(D)J labeling (which requires time-consuming identification of recombination events), we have developed a fast VJ FINDER tool that performs only V and J labeling (Supplemental Material "B. VJ Finder algorithm").

After filtering contaminated reads, the remaining reads are realigned to the starting position of their corresponding V gene. We further define the *distance* between two reads (of possibly different lengths) as the *distance* between the shorter read and the prefix of the longer read of the same length. Two reads are called *similar* if the distance between them does not exceed a *distance threshold* $\tau$. IGREC constructs the *Hamming graph* on the set of all reads as vertices and connects two vertices (reads) in the Hamming graph by an edge if they are similar (Yang et al. 2010; Medvedev et al. 2011; Safonova et al. 2015).

Analysis of the connected components of the Hamming graphs of Rep-seq datasets revealed that they typically consist of *dense* (complete or nearly complete) subgraphs connected by very few edges (Safonova et al. 2015). It further revealed that most vertices in each dense subgraph correspond to reads derived from a single antibody (*simple dense* subgraphs) or from similar antibodies differing from each other by a small number of somatic hypermutations (*composite dense* subgraphs). Our goal is to generate a list of dense subgraphs in the Hamming graph and to further break composite dense subgraphs into subgraphs corresponding to single antibodies.

IGREC builds the Hamming graph (HG CONSTRUCTOR module), partitions its vertices into clusters corresponding to dense subgraphs, and further breaks down composite dense subgraphs as described in (Safonova et al. 2015). The HG CONSTRUCTOR algorithm is described below.

**HG Constructor**

The time- and space-efficient construction of large Hamming graphs is a challenging problem that was addressed in HAMMER (Medvedev et al. 2011) and BAYESHAMMER (Nikolenko et al. 2013). However, these approaches construct Hamming graphs on the set of all *k*-mers from reads (for small *k*), rather than the set of all reads as required for solving the repertoire construction problem. IGREPERTOIRECONSTRUCTOR (Safonova et al.. 2015) is the first read-based algorithm for constructing the Hamming graph, but it becomes too slow in the case of large clonally expanded Rep-seq datasets. Below we describe a novel HG CONSTRUCTOR algorithm for efficient construction of the read-based Hamming graph by finding all similar reads in a high-throughput Rep-seq dataset.

Since brute force computing of distance between all strings in a large set of strings is very time consuming, bioinformaticians use various *filtration techniques* to reduce the number of comparisons. A popular filtration approach is based on the fact that two strings of length $L$ differing in at most $\tau$ positions must share at least one $l$-mer of length $L/(\tau + 1)$ (Knuth 1998). Thus, one can index reads by their $l$-mers, find all pairs of reads that share an $l$-mer, and filter out all other pairs of reads (Pevzner and Waterman 1995; Ma et al. 2002). This *filtration* step is followed by a *verification* step to find the distances between all pairs of reads sharing an $l$-mer.

While this filtration approach worked well for genome comparison (Ma et al. 2002) and read mapping (Lin et al. 2008), it becomes prohibitively time-consuming in the case of immunosequencing reads, as different antibodies often share $l$-mers thus reducing the efficiency of the filtration step. Because all antibodies originating from the same V gene often share an $l$-mer from this gene, the filtration approach will be forced to conduct the verification step for nearly all pairs of reads arising from the same V gene. As a result, while this filtration approach was implemented in IGREPERTOIRECONSTRUCTOR and successfully applied to small Rep-seq datasets, analyzing large hypermutated repertoires turned out to be very time-consuming.

**Minimizers**

To address this problem, we have developed a new filtration strategy based on the following idea (Supplemental Material "C. Constructing the Hamming graph"): instead of using all $l$-mers for filtering (for $l = L/(\tau + 1)$), we use some carefully selected $k$-mers (for $k \leq L/(\tau + 1)$), with the goal of removing $k$-mers shared by many antibodies from the filtering process. The minimizers approach is based on the following observation about two strings of length $L$ differing in at most $\tau$ positions: given a set of $\tau + 1$ non-overlapping $k$-mers in the first string, at least one of them appears in the second string. Specifically, we select $\tau + 1$ *rarest and non-overlapping* $k$-mers (called *minimizers*) in each read, find all reads that contain one of the selected minimizers, and further apply the verification step to the found reads (Supplemental Material "C. Constructing the Hamming graph", Section "The minimizers algorithm"). The notion of a rare $k$-mer is defined with respect to its *multiplicity*, i.e., the number of reads containing this $k$-mer. To find all neighbors of a read $R$ in the Hamming graph, we compute the distance between $R$ and all reads containing at least one minimizer of $R$ (IGREC uses the default values $k = 10$ and $\tau = 4$).

The total multiplicity of minimizers for a given read is exactly the number of computations of the distance for this read. Note that even for a small $k$, the described approach provides the exact solution of the Hamming graph construction problem.

The histogram of positions of rare $k$-mers for $k = 10$ and $\tau = 4$ (Supplemental Material "C. Constructing the Hamming graph", Section "The minimizers algorithm") reveals that the minimizers approach often selects $k$-mers from the *complementary determining regions* (*CDRs*), which represent the most diverged parts of the variable regions of antibodies. However, simply selecting three $k$-mers from three CDRs (CDR1, CDR2, and CDR3) for $\tau = 2$ does not result in a good filtration approach since using small values of $\tau$ often results in breaking the connected components of the Hamming graph corresponding to a single antibody. Also, sequencing errors may turn CDRs into unique sequences, even for reads arising from the same antibody.

**Molecular barcoding and amplification artifacts**

As an alternative to computational error correction of Rep-seq datasets, biologists use unique molecular barcoding (Kinde et al. 2011; Kivioja et al. 2011), an experimental approach that allows one to correct most amplification and sequencing errors, thus enabling analysis of low-abundance receptor sequences (Yaari et al. 2013; Laserson et al. 2014; Vollmers et al. 2013; Shugay et al. 2014; Cole et al. 2016; Turchaninova et al. 2016; de Bourcy et al. 2017). This approach is based on attaching a short unique barcode (12–17 nt) to each RNA molecule, so that all amplified copies of this molecule contain the same barcode. Since all reads corresponding to the same RNA molecule contain the same barcode, errors in reads can be corrected by constructing the consensus of all error-prone reads with the same barcode. However, *barcode errors*, *barcode collisions* and *chimeric reads* lead to new computational challenges, as addressed in the MIGEC (Shugay et al. 2014) and PRESTO (Vander Heiden et al. 2014) tools. These challenges are particularly pronounced in the case of *over-amplified* Rep-seq datasets resulting from multiple cycles of amplification that are used for sequencing low-abundance antibodies.

*Barcode errors.* Some reads in barcoded Rep-seq datasets have amplification errors within barcodes (Figure 2). Our analysis revealed that up to 3% of the clusters in the constructed repertoire contain erroneous barcodes in the case of over-amplified Rep-seq datasets. Ignoring barcoding errors leads to an explosion of small clusters that should be merged with larger ones.

*Barcode collisions.* In an ideal experiment, barcodes of length $k$ are randomly and uniformly generated from the set of all $4^k$ $k$-mers. Thus, if barcodes are applied to a set of $n$ immunoglobulin molecules, we expect only $(n \cdot (n-1))/(2 \cdot 4^k)$ pairs of them to share a barcode (barcode collisions). However, since there exist biases in the barcode generation, barcodes are not uniformly sampled from the set of all $k$-mers thus further increasing the number of barcode collisions. While barcoded Rep-seq datasets analyzed in this paper have a rather low number of collisions (typically, less than 10 for a Rep-seq library consisting of a million reads), some barcoded Rep-seq datasets have significantly more collisions. Ignoring barcode collisions may lead to a loss of natural diversity and corruption of the consensus sequences.

*Chimeric reads*. Some Rep-seq sample preparation protocols produce up to ≈ 8% of chimeric reads. Failure to detect chimeric reads reduces the accuracy of the consensus sequences in the constructed repertoire and may lead to artificially inflated diversity. Most chimeric reads (resulting from "gluing" two distinct immunoglobulin molecules during amplification) are filtered out by VJ FINDER since they contain only partial V or J segments. However, some chimeric reads look like real antibody sequences and thus may artificially inflate the diversity of the constructed repertoire. While most chimeric reads form small clusters, simply discarding small clusters does not address the problem since some chimeric reads form clusters with hundreds of reads. While it is difficult to identify chimeric reads in non-barcoded datasets, we can find them using information about barcodes.

pRESTO does not filter chimeric reads and does not correct for errors in the barcodes, which can lead to abundant erroneous barcodes being assigned into distinct clusters. On the other hand, MiGEC (in the default setting) aggressively discards barcodes with a low quality consensus. As the result, repertoires constructed by pRESTO often retain false clusters for barcode consensus sequence generation, while repertoires constructed by MiGEC often have reduced diversity. Below we describe how IGREC addresses these challenges using BARCODEDIGREC tool.

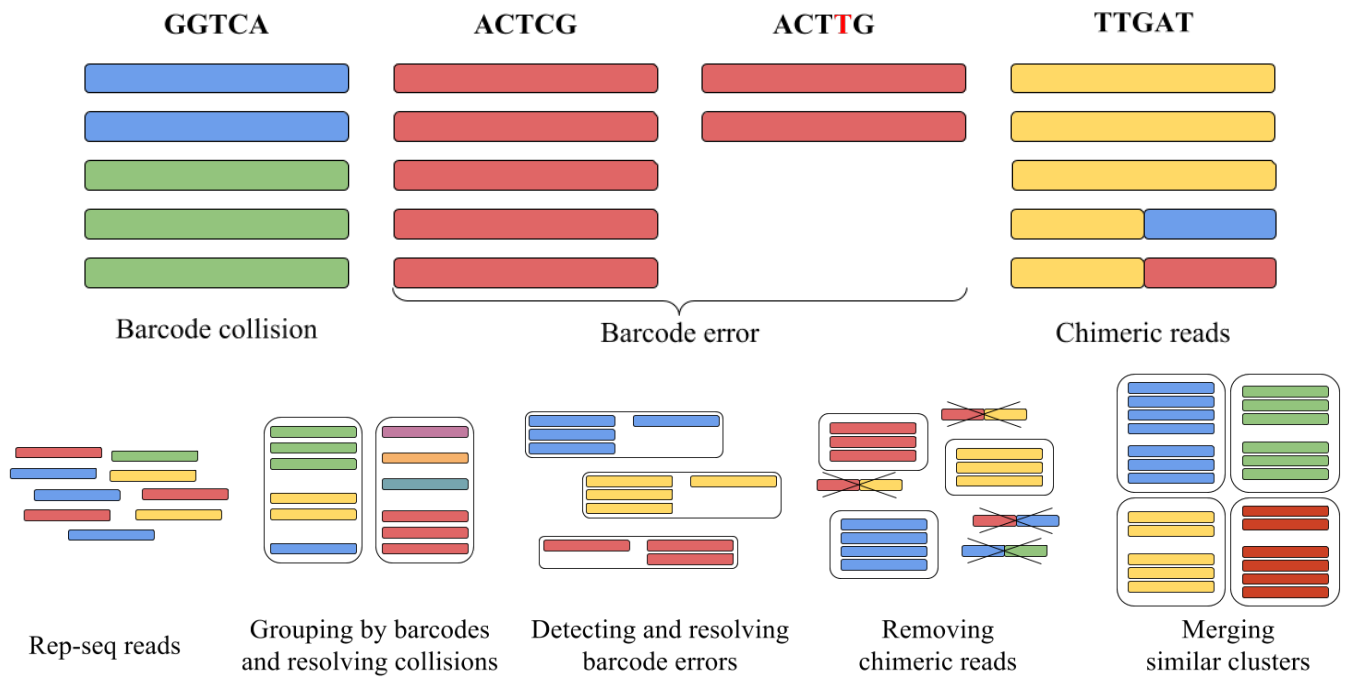

**Figure 2: Barcode/amplification artifacts and the BARCODEDIGREC pipeline.** (Top) Barcode collisions (left), barcode errors (middle), and chimeric reads (right). Each column represents reads with the same barcode shown at the top of the column. Each color represents reads originating from the same RNA molecule. E.g., blue and green reads in the left column share the same barcode GGTCA (barcode collision). Some of the red reads accumulated barcode error during amplification (shown as ACTTG barcode error). The yellow-blue and yellow-red reads are chimeric. (Bottom) Rep-seq reads (1st column) are grouped by barcodes and clustered inside each group to resolve collisions (2nd column). Clusters for similar barcodes are merged if their consensus sequences are similar (3rd column). Chimeric reads are detected and removed (4th column). Finally, the standard IGREC pipeline (for non-barcoded data) is run to merge clusters with similar sequences (5th column).

### BARCODEDIGREC

BARCODEDIGREC constructs clusters using information about barcodes (first four steps in Figure 2 (bottom)) and applies the standard IGREC pipeline to merge similar corrected sequences and to construct the final repertoire (last step in Figure 2 (bottom)). Since IGREC at the final step is applied to highly accurate sequences, we use a small distance threshold for clustering (the default value $\tau_{igrec} = 2$). Supplemental Material "D. Selecting BARCODEDIGREC parameters" describes selection of parameters $\tau_{igrec}$, $\tau_{cluster}$, and $\tau_{umi}$ in BARCODEDIGREC.

BARCODEDIGREC clusters reads sharing the same barcode and computes the consensus for each constructed cluster. We use a large edit distance threshold for clustering to correct most amplification and sequencing errors and, at the same time, separate barcode collisions. We use the edit distance (rather than Hamming distance), because barcoded Rep-seq dataset have amplification errors that include both indels and mismatches. Since our analysis of various Rep-seq datasets revealed that the vast majority of randomly selected antibodies have more than 20 differences, we set the default value $\tau_{cluster} = 20$. To correct barcode errors, we construct the Hamming graph on the barcodes with the default value $\tau_{umi} = 1$ (step 2 in Figure 2 (bottom)) and merge clusters with similar consensus sequences for barcodes from the same connected components in the Hamming graph. If there are several clusters for a single connected component of a Hamming graph on barcodes, we also check them for chimerism. We classify a cluster as chimeric if its suffix coincides with a suffix of a consensus string from a different barcode.

**Constructing reference repertoires**

Tools like QUAST (Gurevich et al. 2013) benchmark various assembly algorithms against the *reference genomes*. However, in many areas of genomics, reference genomes remain unknown; e.g., METAQUAST (Mikheenko et al. 2016) evaluates metagenomics assemblies without references. Similar to metagenomics, reference antibody repertoires for Rep-seq datasets are not available.

To construct reference repertoires, we use simulated Rep-seq datasets generated by IGSIMULATOR (Safonova et al. 2015a). However, simulated datasets do not provide comprehensive benchmarking since IGSIMULATOR does not adequately model some intricate properties of real Rep-seq datasets. We thus complement simulated datasets by the repertoires derived from barcoded Rep-seq data in a blind analysis; i.e., without using the barcoding information.

**Sensitivity and precision of a repertoire**

A cluster in the constructed repertoire is *correct* if its consensus sequence is present in the reference repertoire (see Supplemental Material "E. Abundance analysis with IGQUAST" to see how IGQUAST defines matches between sequences in the constructed and reference repertoires). Given the constructed and the reference repertoires of sizes $N_{con}$ and $N_{ref}$, we define the *sensitivity* as $N_{corr}/N_{ref}$ and the *precision* as $N_{corr}/N_{con}$, where $N_{corr}$ is the number of correct clusters (Figure 3a). Low sensitivity often results from *overcorrection:* combining multiple clusters from the reference repertoire into a single cluster in the constructed repertoire. Low precision often results from *undercorrection:* splitting a single cluster from the reference repertoire into multiple clusters in the constructed repertoire.

Since cluster abundances in a repertoire are well approximated by the power law distribution, there are typically a small number of large clusters and many small clusters (Weinstein et al. 2009). While the consensus sequences of large clusters are typically accurate, the consensus sequences of small clusters often have errors (even in the case of correctly constructed clusters). E.g., the consensus sequence of a cluster of size 2 is often corrupted if the reads contributing to this cluster have errors. Thus, small clusters should be excluded from computing sensitivity and precision since these parameters are computed for exact matches between consensus sequences in the constructed and reference repertoires. Also, the constructed repertoires are often more diverse than the reference repertoire since clusters in the reference are often broken into several clusters in the constructed repertoire due to sequencing and amplification errors. Typically, each large cluster in the reference repertoire corresponds to one large cluster with a correct consensus sequence and a number of small clusters with erroneous consensus sequences (Figure 3(b)).

To exclude small clusters from computing sensitivity and precision, we consider only large clusters of size at least $minsize_{ref}$ in the reference repertoire and of size at least $minsize_{con}$ in the constructed repertoire (the default value $minsize_{ref} = 5$). The optimal value of $minsize_{con}$ is selected based on analysis of the *sensitivity-precision plot* constructed for various values of $minsize_{ref}$ (Figure 4 (a)).

IGQUAST also computes the *abundance plot* by comparing the abundances of all correct clusters in the constructed and reference repertoires (Figure 4 (b)). Since clusters in the reference repertoires are often broken into several clusters in the constructed repertoire, the abundances of clusters in the constructed repertoires are typically lower than in the reference repertoire. IGQUAST uses the sensitivity-precision and the abundance plots for benchmarking various repertoire construction tools (see Supplemental Material "E. Abundance analysis with IGQUAST").

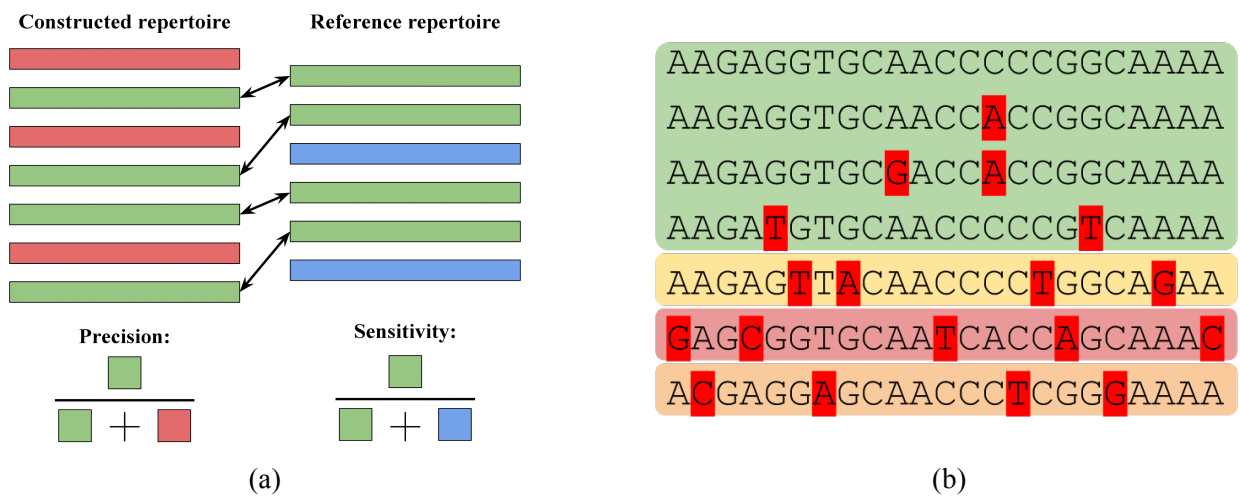

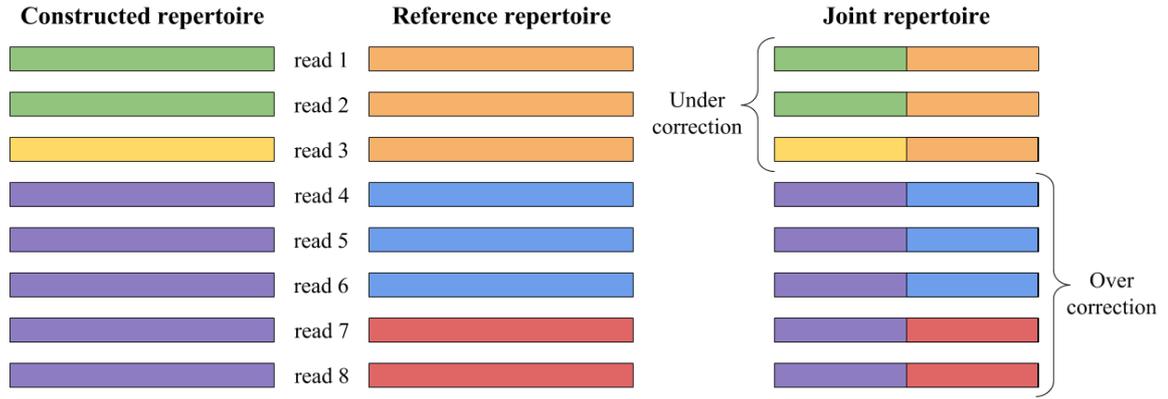

(c)

**Figure 3: Precision, sensitivity, and joint repertoires.** (a) Precision is the fraction of correct (green) clusters among all clusters in the constructed repertoire (green and red) equal to 4/7. Sensitivity is the fraction of correct (green) clusters among all clusters in the reference repertoire (green and blue) equal to 4/6. (b) Each cluster in the reference repertoire is typically broken into multiple clusters (shown by four different colors) in the constructed repertoire. (c) The constructed (left column), the reference (middle column), and the joint (right column) repertoires represent various partitions of reads into clusters, the joint repertoire consists of four cluster while the constructed and reference repertoires consisting of three clusters each. The orange cluster in the reference repertoire is under corrected since its reads belong to green and yellow clusters in the constructed repertoire. For this cluster, the largest cluster in the joint repertoire has size 2, implying that its impurity is equal to 1/3. The constructed violet cluster is overcorrected since its reads include both blue and red reference clusters. Note that clusters in the constructed and references repertoires do not necessarily include each other.

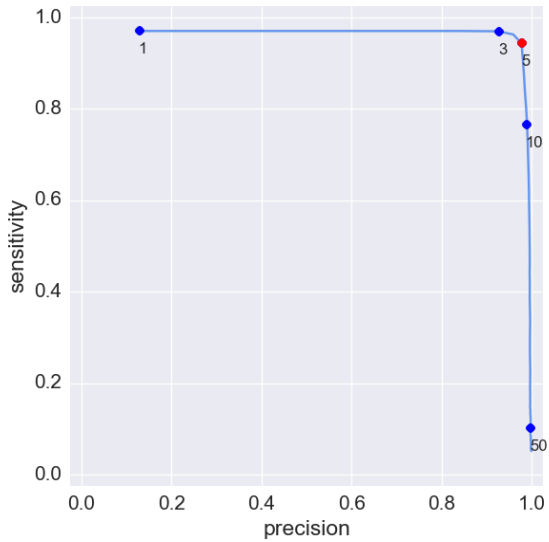

(a)

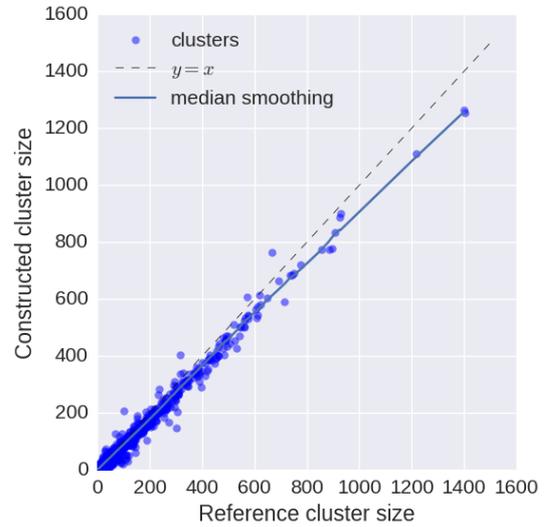

(b)

**Figure 4: Sensitivity-precision and abundance plots.** (a) Each point on the sensitivity-precision plot represents the precision (x-axis) and the sensitivity (y-axis) for the default value $minsize_{ref} = 5$, and a specific value of $minsize_{con}$. The red point corresponds to the $minsize_{ref}$ value that maximizes the sum of sensitivity and precision. (b) Each correct cluster is represented by a point in the abundance plot whose $x$ and y coordinates correspond to abundances of this cluster in the reference and the constructed repertoires, respectively. The color of each point $(x, y)$ represents the number of clusters with abundances in the reference and constructed repertoires equal to $x$ and $y$, respectively (the deeper the color, the larger the number of clusters). The $x = y$ line corresponds to the correctly reconstructed abundances. The blue line shows the median for the abundances of the constructed clusters. Data is shown for the REAL dataset described in the Results section.

**Detecting overcorrected clusters**

While IGREC typically undercorrects clusters in the constructed repertoires, it may also overcorrect highly similar clusters (e.g., clusters specific to the same antigen) that belong to the same clonal lineage. Thus, detection of overcorrected clusters is important for accurate reconstruction of clonal diversity.

Since the consensus sequences of overcorrected clusters are often corrupted, it is difficult to detect overcorrection using *exact matching* of sequences in the constructed repertoire against sequences in the reference repertoire (as was done for computing the sensitivity and precision). To detect overcorrected clusters, we constructed a *joint repertoire* (for constructed and reference repertoires) as follows. For each cluster $A$ in the constructed repertoire that shares reads with a cluster $B$ in the reference repertoire, we define the cluster $A \cap B$ in the joint repertoire. Each cluster $A$ in the constructed (reference) repertoire is the union of clusters in the joint repertoire. We denote the largest cluster in this union as $A_1$ and define the *impurity* of a cluster $A$ as $1 - |A_1|/|A|$ (Figure 3c). IGQUAST reports clusters with non-zero impurity as overcorrected clusters (Supplemental Material "F. Detection overcorrection with IGQUAST", Section "Reference-based detection of overcorrection"). IGQUAST can also detect overcorrection even in the case when the reference repertoire is unknown (see Supplemental Material "F. Detection overcorrection with IGQUAST", Section "Reference-free detection of overcorrection" for details).

**Reference-free analysis of overcorrection**

Analysis of overcorrected clusters can be extended to cases where the reference repertoire is unknown. IGQUAST aligns all reads from each large cluster against the consensus sequence of the cluster and analyzes all columns in the resulting multiple alignment. For each column, it computes the *discordance*, which is the fraction of reads corresponding to the second most abundant nucleotide in this column. The *discordance* of a cluster in the constructed repertoire is defined as the maximal value of discordances among all columns. Gluing multiple similar antibodies into a single cluster results in high discordance and reveals overcorrection. IGQUAST reports clusters with large discordance as potentially overcorrected (see Supplemental Material "F. Detection overcorrection with IGQUAST", Section "Reference-free detection of overcorrection" for details).

Our analysis revealed that in the case of repertoires containing small clonal trees (e.g., bulk repertoire of peripheral blood of a healthy individual), the number of overcorrected clusters is small for all analyzed tools and the quality of the constructed repertoires (sensitivity) hardly changes if we split the overcorrected clusters. However, in cases of repertoires with large clonal trees (e.g., repertoires of antibody-secreting cells after vaccination), splitting overcorrected clusters significantly improves the quality of a constructed repertoire.

**Analyzing diversity of adaptive immune repertoires**

In addition to reporting SHM statistics and analyzing nucleotide and amino acid content of CDRs, our IgDIVERSITYANALYZER tool complements IMGT/HIGHV-QUEST, MIXCR, VDJTOOLS, and ALAKAZAM tools by reporting the *Simpson index* and the *clonal Simpson index* of the repertoire (see Supplemental Material "G. IgDIVERSITYANALYZER" for details).

The Simpson index, originally proposed for analyzing diversity of populations (Simpson 1949), is computed as the probability that two randomly selected individuals belong to the same population. Given a probability distribution $(p_1, \ldots, p_N)$, where $p_i$ represents the fraction of the *i*-th population, the Simpson index is defined as $SI = \sum_1^n p_i^2$. IGDIVERSITYANALYZER uses *CDR3 abundances* (the sum of abundances of clusters with the same CDR3) and normalizes them to generate the probability distribution for computing the Simpson index. Diverse repertoires are characterized by low Simpson index, while repertoires dominated by a few abundant receptor sequences are characterized by high Simpson index.

IgDIVERSITYANALYZER also computes the *clonal Simpson index* (*CSI*) to reveal similarities between CDR3s in a repertoire. It constructs the Hamming graph on CDR3 sequences with a small distance threshold (the default value $\tau = 3$) and computes abundance of each connected component as a sum of its CDR3 abundances. The CSI is defined as the Simpson index on abundances of the connected components in the constructed Hamming graph. The CSI can be interpreted as a probability that two randomly selected antibody sequences belong to the same clonal lineage. The higher the ratio *CSI / SI*, the higher the clonal diversity of a repertoire, as compared to its recombination diversity.

# Results

**Benchmarking approaches**

We benchmarked IGREC, MIXCR and PRESTO using:

- reference repertoires and corresponding Rep-seq libraries simulated by IGSIMULATOR (Safonova et al. 2015a),

- reference repertoires derived from barcoded Rep-seq repertoires, but applied these tools in blind mode; i.e., without using information about barcodes,

- a reference-free approach for evaluating repertoires applied to Rep-seq datasets from various B cell subtypes. Since these datasets have widely varying levels of diversity, we analyzed how various repertoire construction tools preserve natural diversity.

**Datasets**

We benchmarked IGREC, BARCODEDIGREC, MIXCR, MIGEC, and PRESTO on simulated, synthetic, and real Rep-seq datasets. Characteristics of these datasets are summarized in
Table *1*. PBM, AS-, and AS+ Rep-seq datasets were described in Ellebedy et al. 2016).

- *SIMULATED* datasets were generated by IGSIMULATOR (Safonova et al. 2015a) with the following parameters: *number of base sequences* = 100, *expected number of mutated sequences* = 1000 and *expected repertoire size* = 5000. While the SIMULATED datasets do not adequately reflect some features of real antibody repertoires, it provides a way to evaluate the constructed repertoires for various error rates.

- IGSIMULATOR uses the ART ILLUMINA read simulator (Huang et al. 2012) to generate a sequencing library along with a simulated repertoire. However, this and other read simulators create reads with typical error rates in Illumina reads (~1% of incorrect bases per read) and do not account for elevated rates of amplification errors in Rep-seq datasets. On the other hand, some Rep-seq datasets have error rates as low as 0.25% (Safonova et al. 2015). We thus simulated Rep-seq datasets with both low and high error rates by implanting random mismatches into randomly chosen positions in sequences from the reference repertoire. We selected 12 values for the average number of errors in reads ranging from unrealistically low to rather high (0, 0.0625, 0.125, 0.25, 0.375, 0.5, 0.75, 1, 1.25, 1.5, 1.75, and 2) and simulated a Rep-seq library, where the number of mismatches in reads followed the Poisson distribution with the parameter $\lambda$ equal to the average number of mismatches in the simulated Rep-seq library. We refer to them as SIMULATED SIMPLE datasets because they often contain error-free reads making it easier to reconstruct a repertoire. As a result, for each input repertoire we simulated 12 SIMULATED SIMPLE and 12 SYNTHETIC SIMPLE datasets. We also simulated 12 more difficult-to-analyze SIMULATED COMPLEX datasets (modeling highly corrupted clusters) where each input read contains at least one sequencing error.

- The *SYNTHETIC* dataset was generated using an antibody repertoire generated by Juno Therapeutics based on a barcoded Rep-seq dataset. The SYNTHETIC dataset is a challenging test since it represents a clonally expanded post-vaccination repertoire containing extensively hypermutated antibodies. Both SIMULATED and SYNTHETIC repertoires were used as a base for simulation of Rep-seq libraries with various error rates.

- The *REAL* dataset represents a publicly available barcoded Rep-seq data of bulk unsorted B cells from peripheral blood of a healthy donor generated at the Institute for Bioorganic Chemistry in Moscow, Russia. For benchmarking purposes, we compared repertoires constructed by IGREC, MIXCR, and PRESTO tools in blind mode with the reference repertoire obtained using BARCODEDIGREC. The REAL dataset does not contain highly abundant clonal families, but reflects the complexity of real data.

- The *PBM dataset* was derived from peripheral blood mononuclear B cells. The PBM cells were taken right after flu vaccination (day 0), while antibody secreting cells were sequenced and sorted at the moment of the highest immune response to flu vaccine (day 7). The PBM cells include various B cell subtypes (naïve, memory, and plasma) and thus, are characterized by diverse recombination events and large variations in antibody abundances.

- The *AS– dataset* was derived from antibody secreting cells negative to hemagglutinin. In contrast to the PBM dataset, the antibody secreting cells generate highly abundant antibodies. The B cells from the AS+ dataset share specificity to hemagglutinin and, thus, are expected to form abundant clonal lineages

- The *AS+ dataset* was derived from antibody secreting cells positive to hemagglutinin. Since the B cells from the AS– dataset showed negative specificity to hemagglutinin, we expect that the most abundant clonal lineages from this dataset are, nevertheless, less abundant than the most abundant clonal lineages from the AS+ dataset.

**Analyzing Rep-seq datasets of sorted B cells**

The specific features of sorted cells allow one to evaluate the results of various repertoire construction tools. E.g., highly similar antibodies in a PBM repertoire is a sign of undercorrection, while a lack of natural variations in an AS+ repertoire is a sign of overcorrection or sample preparation artifacts.

We characterize each repertoire by its recombination (two parameters), expression (three parameters), and clonality levels (two parameters), using the following statistics (

Table *1*):

- *recombination level*: number of unique CDR3s and VJ pairs;
- *expression level*: max cluster size, max CDR3 abundance, Simpson index (*SI*);
- *clonality level*: clonal Simpson index (*CSI*), the ratio *CSI / SI*.

We analyzed PBM, AS–, and AS+ datasets using IGREC and selected large clusters in the constructed repertoires ($minsize_{con} = 5$) to minimize the effect of amplification artifacts on our analysis. We extracted the CDR3 sequence for each constructed cluster using IGDIVERSITYANALYZER and grouped together identical CDR3s. Since CDR3 is the most divergent part of antibodies, the set of all CDR3s is a good (albeit insufficient) representation of the diversity of the entire repertoire. For the PBM repertoire, we expect to observe many dissimilar CDR3s corresponding to unrelated V(D)J recombinations; while for the AS+ repertoire we expect to observe large groups of similar CDR3s that came from the same antibody lineage and differ by SHMs. We also expect that various metrics for the AS– repertoire have smaller values than those of the AS+ repertoire, but larger values than for the PBM repertoire.

Table *1* (middle) illustrates that the IGREC results for the PBM, AS– and AS+ repertoires are consistent with expectations summarized in

Table *1* (top):

- *Recombination levels.* The PBM repertoire contains more CDR3s (6255 vs 5008 and 949), and more VJ pairs (885 vs 537 and 156) than the AS– and AS+ repertoires, respectively.

- *Expression levels.* While the PBM repertoire contains some highly abundant clusters (the largest is formed by 3834 reads), its CDR3s abundances are rather low (max CDR3 abundance is 20) and its Simpson index is only 0.00024. The AS– and AS+ repertoires contain more abundant clusters, many of which share CDR3. The maximal cluster for the AS+ repertoire is very large (47,750 reads), which reflects a clonal expansion process.

- *Clonality level.* The ratio of the Simpson index and the clonal Simpson index revealed that the PBM repertoire contains very limited number of clonal lineages ($CSI / SI$ = 1.36). In contrast, the AS– and AS+ repertoires have large clonal lineages ($CSI / SI$ is 7.15 and 11.96 for AS– and AS+, respectively).

**Benchmarking on the SIMULATED datasets**

We ran MIXCR, PRESTO and IGREC on each simulated dataset (see Supplemental Material "I. Benchmarking parameters") and selected the parameter $minsize_{con}$ that maximizes the sum of sensitivity and precision for each constructed repertoire. Figure 5 (a,b) shows the sensitivity and precision plots for the SIMULATED SIMPLE datasets and illustrates that all tools deteriorate with the increase in the error rate. However, IGREC reconstructs 73% of reference clusters, even in the case of high number of errors equal to 2, while the corresponding values of sensitivity for MIXCR and PRESTO fall to 20%.

We also benchmarked IGREC, MIXCR, and PRESTO for typical values of error rates in real datasets corresponding to 0.5 and 1 errors per read on average (Figure 6 (a, b)). IGREC improved on MIXCR and PRESTO for 0.5 and 1 errors, except for a small region in the sensitivity-precision plot for 0.5 errors (corresponding to high precision). PRESTO improved on MIXCR in the case of 0.5 errors, while MIXCR improved on PRESTO in the case of 1 error.

In contrast to IGREC, MIXCR and PRESTO failed to construct correct cluster sequences for the SIMULATED COMPLEX datasets. The repertoires constructed by IGREC for the COMPLEX SIMULATED dataset are even more accurate than repertoires constructed from SIMULATED SIMPLE datasets.

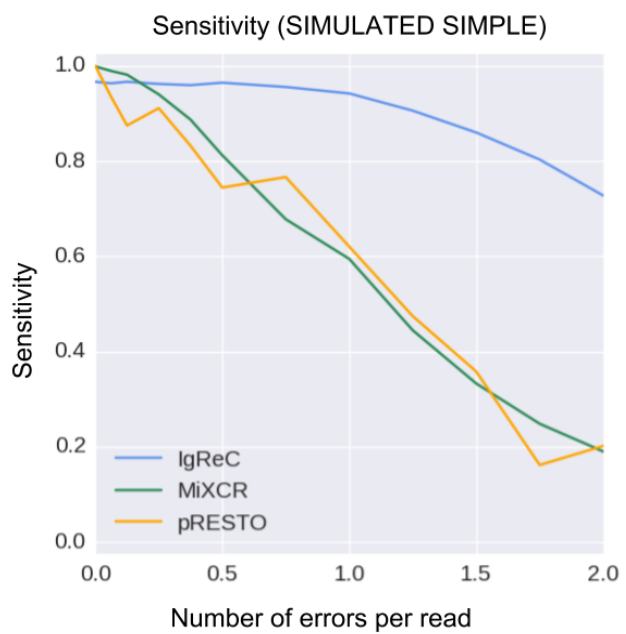
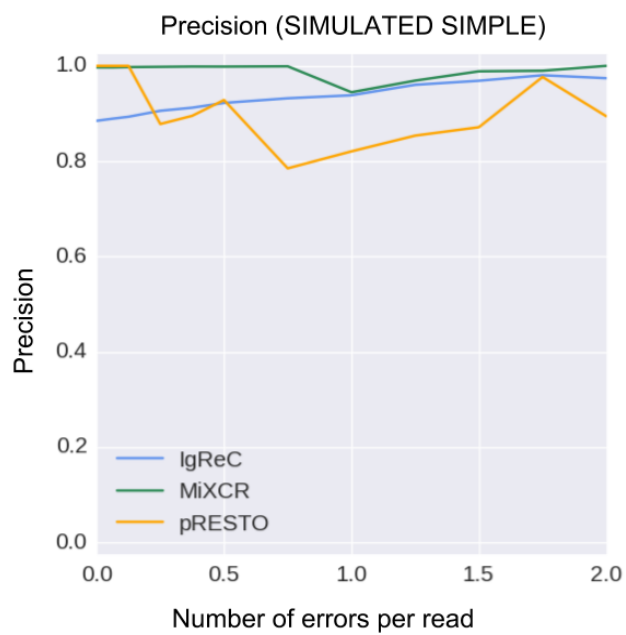

(a)                  (b)

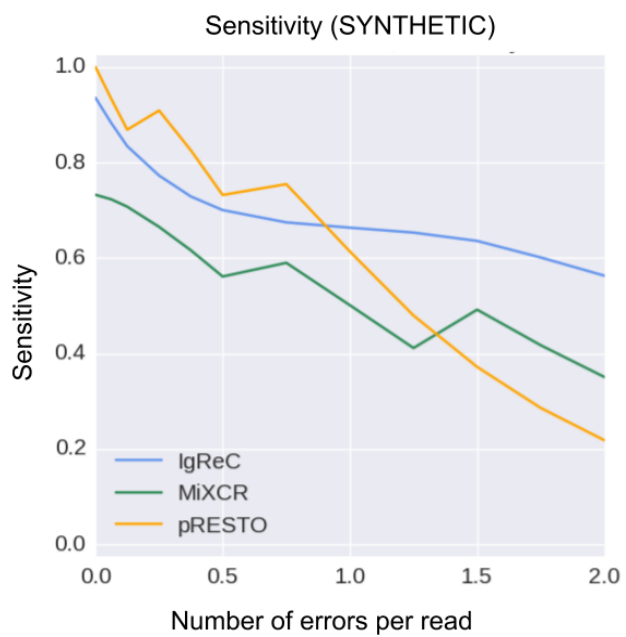
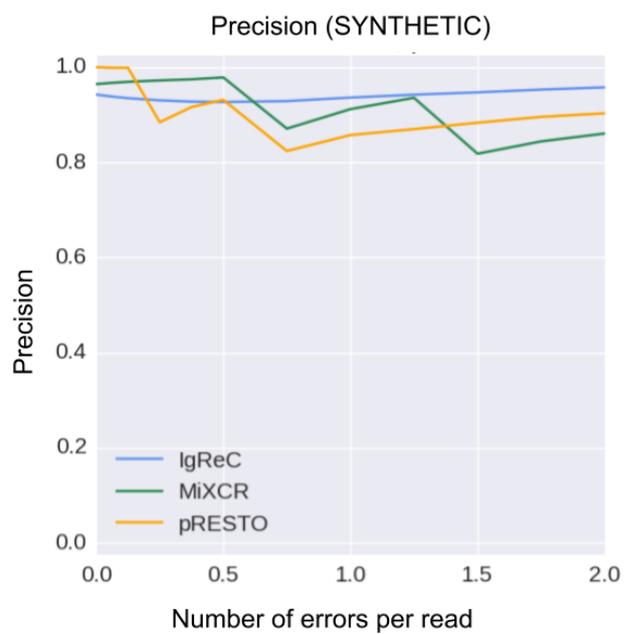

(c)                  (d)

**Figure 5: Sensitivity and precision for various error rates for SIMULATED SIMPLE (a, b) and SYNTHETIC SIMPLE (c, d) datasets.** The plots are constructed for IGREC (blue), MIXCR (green), and PRESTO (orange). For IGREC, the optimal value of $minsize_{con}$ is equal to 5 for all error rates, while for MIXCR and PRESTO, the optimal value of $minsize_{con}$ varies from 2 to 5, depending on error rates.

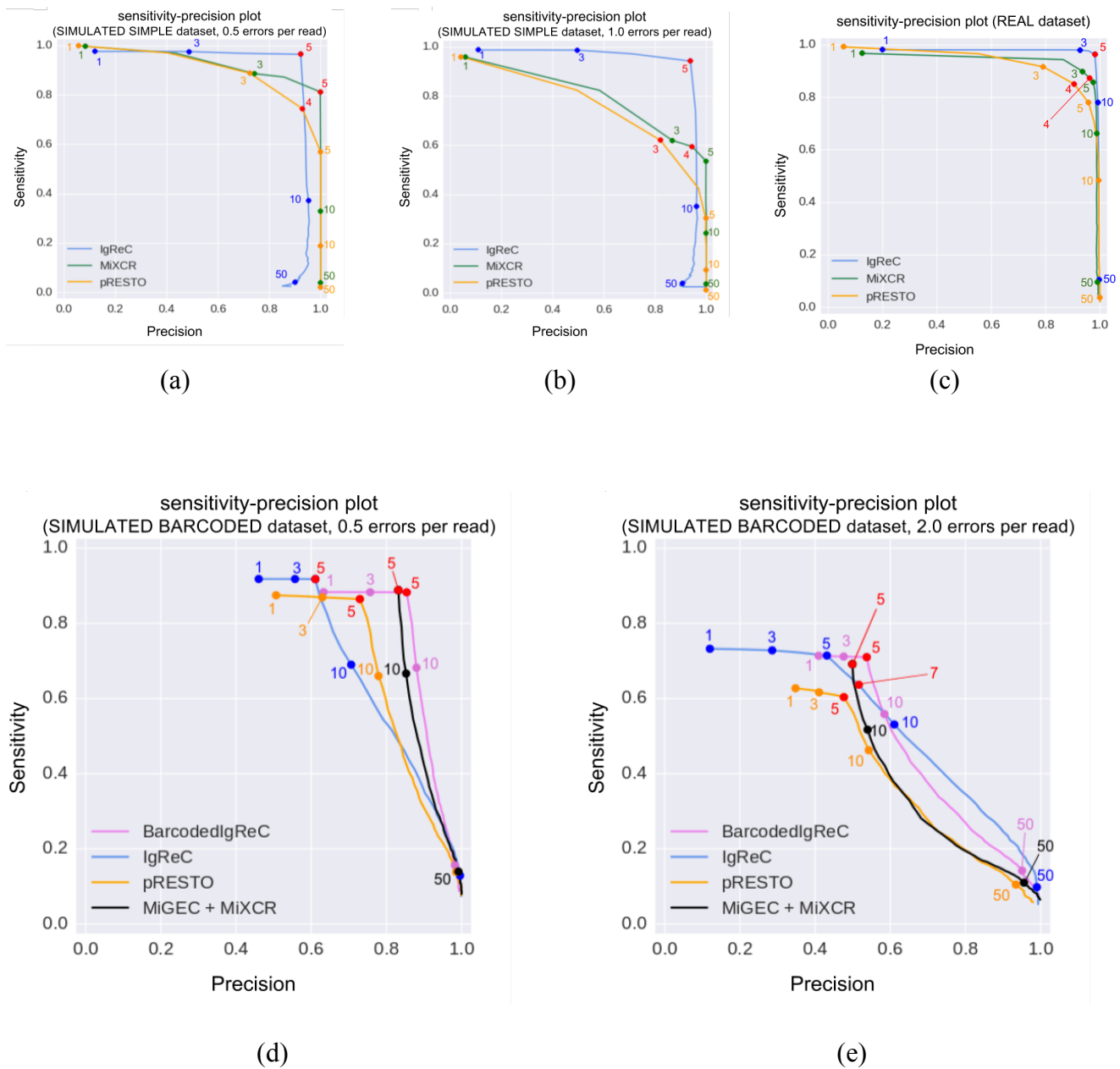

**Figure 6: The sensitivity-precision plots for repertoires constructed by IGREC, BARCODEDIGREC, PRESTO, MIXCR, and MIGEC+MIXCR.** (Top) The sensitivity-precision plots for the SIMULATED SIMPLE (a, b) and the REAL (c) datasets. (Bottom) The sensitivity-precision plots for repertoires constructed by BARCODEDIGREC, IGREC, PRESTO, and MIGEC+MIXCR for SIMULATED BARCODED dataset with low (d) and medium (e) error rates. Red points correspond to the $minsize_{con}$ value maximizing the sum of sensitivity and precision.

**Benchmarking on the SYNTHETIC dataset**

Figure 5 (c,d) illustrates that the sensitivity values for the SYNTHETIC dataset are lower as compared to the SIMULATED SIMPLE datasets, reflecting the fact that there exist many similar clusters in the reference repertoire for the SIMULATED SIMPLE dataset. Since the SYNTHETIC dataset is very complex, we limited benchmarking on this dataset to simple libraries only.

All tools resulted in high precision on the SYNTHETIC dataset for all values of error rates (Figure 5 (c)). However, MIXCR and PRESTO showed significantly lower sensitivity than IGREC for high error rates (Figure 5 (d)). In the case of low number of errors (<0.8), PRESTO showed higher sensitivity than IGREC. In general, IGREC maintained high values of sensitivity for the entire range of error rates (93% for error rate 0 and 56% for error rate 2).

**Benchmarking on the REAL dataset**

Figure 6 (c) illustrates that all tools demonstrate high sensitivity and precision on the REAL dataset. IGREC slightly improves on both MIXCR and PRESTO, and MIXCR slightly improves on PRESTO. It turned out that IGREC works well with error-prone and over-amplified data: its sensitivity (98%) and precision (96%) for the REAL dataset are even higher than the corresponding values for the SIMULATED dataset with 0 error rate (sensitivity 97%, precision 88%).

**Benchmarking on the barcoded datasets**

Using barcodes improves the performance of BARCODEDIGREC, MIGEC and PRESTO as compared to applying these tools in blind mode; i.e., ignoring barcodes. Therefore, repertoires constructed using information about barcodes represent excellent references for repertoires constructed in blind mode. However, it is important to evaluate how repertoire construction tools perform on barcoded datasets to measure how close the resulting repertoires are to the "ideal" repertoires.

To address this problem, we extended IGSIMULATOR by adding a procedure for simulating realistic barcoded Rep-seq datasets featuring barcode errors, barcode collisions, and chimeric reads (Supplemental Material "K. Extending IgSimulator to barcoded Rep-seq datasets"). Our analysis revealed that reads in barcoded Rep-seq datasets contain 2-3 errors on average. Thus, we simulated low (0.5 errors per read) and medium (2 errors per read) error rates respectively, which resulted in two SIMULATED BARCODED libraries with 478,250 and 481,245 reads. We further compared repertoires constructed by IGREC (in blind mode), BARCODEDIGREC, MIGEC + MIXCR and PRESTO. We note that MiGEC was used for clustering of reads with the same barcodes and correction of barcode errors. To construct the full-length repertoire, we launched MiXCR on the consensus sequences reported by MiGEC as proposed in (Turchaninova et al. 2016).

All tools demonstrated high sensitivity for the library with low error rate (Figure 6 (d)), but BARCODEDIGREC and MIGEC+MIXCR outperformed IGREC and PRESTO with respect to precision. Specifically, BARCODEDIGREC and MIGEC+MIXCR correctly reconstructed 85% and 83% of large clusters (more than 5 reads), respectively (compared to 61% for IGREC and 73% for PRESTO). Surprisingly, IgReC launched in blind mode on a library with high error rates improved on the specialized tools utilizing barcoding information (Figure 6 (e)). See Supplemental Material "L. Benchmarking on barcoded Rep-seq datasets with various error rates."

## Discussion

We presented the IGREC algorithm for antibody repertoire construction and benchmarked it against the state-of-the-art tools MIGEC, MIXCR and PRESTO. Although it is difficult to draw a fair comparison of these tools (they all have slightly different goals), comprehensive assessment of various tools is crucial for all areas of modern genomics and immunogenomics is not an exception. Since such benchmarking effort in immunogenomics is still missing, we developed the IGQUAST tool for the quality assessment of the constructed repertoires. IGQUAST enabled the first benchmarking study of various repertoire reconstruction approaches across diverse Rep-seq datasets. We also proposed a new reference-free approach to quality assessment of the antibody repertoires using Rep-seq datasets from sorted B cells with varying diversity and expression levels. As a result, repertoires constructed by IGREC reflect the biological features and the diversity of these repertoires. Although IgReC, MiXCR, and pRESTO showed different results in terms of sensitivity and precision, our analysis revealed that all tools accurately reflect important biological features of Rep-seq data.

In difference from other repertoire construction tools, IGREC reconstructs repertoires based on the concept of the Hamming graphs (Medvedev et al. 2011; Nikolenko et al. 2013), the workhorse of the error-correction approach in the leading genome assembler SPAdes (Bankevich et al. 2012). To address the computational bottlenecks of constructing the immunosequencing Hamming graph, we developed a minimizers approach that resulted in a three orders of magnitude speed-up as compared to IGREPERTOIRECONSTRUCTOR (Supplemental Material "M. Benchmarking IgReC vs IgRepertoireConstructor"). As a result, IGREC analyzes large highly hypermutated Rep-seq datasets in minutes compared to the days required by IGREPERTOIRECONSTRUCTOR (Safonova et al. 2015). The IGREC algorithm is currently being extended to analyzing TCR repertoires.

Our benchmarking revealed that there is still no single repertoire construction tool that works better than other tools across the diverse types of Rep-seq datasets. For example, repertoires constructed by IGREC featured better sensitivity than repertoires constructed by MIXCR and PRESTO, but had lower precision in some cases. However, as our benchmarking demonstrated, IGREC is currently a tool of choice for analyzing hypermutated repertoires. In this case, it is more important to extract detailed information about expanded clonal lineages, while erroneous clusters can be filtered at a later stage using the clonal analysis (Galson et al. 2016). On the other hand, high precision is critical for diversity analysis of repertoires from healthy individuals. Since such repertoires typically do not contain expanded clonal lineages, large erroneous clusters may negatively affect the estimates of the diversity of the constructed repertoire.

We compared BARCODEDIGREC, MIGEC and PRESTO tools that use barcode information to construct repertoires. While barcoding significantly simplifies repertoire reconstruction and allows one to recover low-abundance antibodies from extensively amplified Rep-seq libraries, it also leads to various experimental artifacts that may trigger errors in the reconstructed antibody repertoires. Despite the fact that the mitigation of barcoding artifacts in genome sequencing is extensively described in literature (Hamady et al. 2008; Krishnan et al. 2011; Buschmann and Bystrykh 2013), the efforts to address these artifacts in immunogenomics are still in their infancy. Our attempts to address them led to the surprising conclusion that in many immunogenomics studies barcoding may not be necessary since our computational approach to error correction of immunosequencing data (Hamming graphs) ends up being nearly as powerful as the experimental approach to error correction (barcoding).

Our benchmarking revealed that while various repertoire construction tools result in high sensitivity, their precision varies and becomes rather low in the case of Rep-seq datasets with many amplification errors. Surprisingly, repertoires reported by IgReC tool (in blind mode without using barcoding information) on barcoded datasets improved on the repertoires constructed by the specialized tools that use barcoding information. This finding suggests that advanced error correction algorithms may alleviate the need to generate barcoded repertoires, thus reducing experimental effort. While barcoded Rep-seq datasets may still be needed for specialized studies aimed at generating extremely accurate antibody repertoires or quantification (section "Abundance analysis on barcoded datasets using barcode counting"), we believe that repertoires constructed by IgReC from non-barcoded Rep-seq datasets are well-suitable for most existing immunogenomics applications, e.g., clonal analysis or immunoproteogenomics.

## Availability

## Acknowledgement

This project is supported by Russian Science Foundation (grant No 14-50-00069). Immunoglobulin library preparation was supported by Russian Science Foundation grant No 14-14-00533 (to Maria Turchaninova). We are indebted to Dmitry Bolotin, Mikhail Shugay, Dmitriy Chudakov, Jason Vander Heiden, and Steven Kleinstein for productive discussions and assistance in benchmarking MiXCR, MiGEC, and pRESTO tools. We thank Sonya Timberlake for providing us with the SYNTHETIC dataset and Anton Bankevich for insightful comments and help in preparation of this paper.

## Disclosure declaration

Authors have no conflicts to report.

|  | **PBM** | **AS–** | **AS+** |
|---|---|---|---|
| **Recombination level** | High | Medium | Low |
| **Expression level** | Medium | High | Very high |
| **Clonality level** | Low | Medium | High |

|  | **PBM** | **AS–** | **AS+** |
|---|---|---|---|
| *# reads* | 446,393 | 285,893 | 355,405 |
| *# large clusters* | 8254 | 7325 | 3654 |
| *# reads in large clusters* | 188,329 | 262,738 | 315,846 |
| *# unique CDR3s* | 6255 (76%) | 5008 (63%) | 949 (26%) |
| *# VJ pairs* | 885 (22%) | 537 (13%) | 156 (4%) |
| *max cluster size* | 3834 (2%) | 8347 (3%) | 47,750 (15%) |
| *max CDR3 abundance* | 20 (0.24%) | 140 (2%) | 678 (19%) |
| *SI* | 0.00024 | 0.00088 | 0.04183 |
| *CSI* | 0.00033 | 0.00626 | 0.50037 |
| *CSI / SI* | 1.36 | 7.15 | 11.96 |

|  | **SIMULATED** | **SYNTHETIC** | **REAL** |
|---|---|---|---|
| *# reads* | 44,094 | 793,227 | 500,757 |
| *# large clusters* | 1549 | 31,297 | 15,391 |
| *# reads in large clusters* | 26,276 | 547,855 | 490,997 |
| *# unique CDR3s* | 1217 (78%) | 20,452 (65%) | 14,822 (96%) |
| *# VJ pairs* | 315 (8%) | 1106 (27%) | 691 (17%) |
| *max cluster size* | 2292 (9%) | 6429 (1%) | 1403 (0.3%) |
| *max CDR3 abundance* | 41 (3%) | 563 (2%) | 17 (0.1%) |
| *SI* | 0.0023 | 0.0003 | 0.00007 |
| *CSI* | 0.024 | 0.0004 | 0.0001 |
| *CSI / SI* | 10.43 | 1.33 | 1.39 |

**Table 1. Diversity metrics for repertoires constructed by IGREC.** (Top) Properties of PBM, AS– and AS+ cells. Rows corresponding to recombination, expression, and clonality levels are colored by various shades of blue, green, and red, respectively. High (low) levels of these metrics correspond to dark (light) shades. (*Middle*) Diversity metrics for repertoires constructed by IGREC. The shown percentages represent percentage of clusters with unique CDR3s among all large clusters, percentage of detected VJ pairs among all possible VJ pairs, percentage of reads in the largest cluster among all reads in large clusters, and percentage of reads with the most abundant CDR3 among all reads in large clusters. The diversity metrics of the repertoires constructed by MIXCR and PRESTO are given in Supplemental Material "H. Diversity analysis of repertoires constructed by MiXCR and pRESTO." (Bottom) Diversity metrics for SIMULATED, SYNTHETIC, and REAL repertoires.

# Reconstructing antibody repertoires
# from error-prone immunosequencing datasets
# (Supplemental Material)

Alexander Shlemov[1,*], Sergey Bankevich[1,*], Andrey Bzikadze[1,2],
Maria A. Turchaninova[3], Yana Safonova[1,**] and Pavel A. Pevzner[1,4]

[1]Center for Algorithmic Biotechnology, Institute for Translational Biomedicine,
St. Petersburg University, St. Petersburg, Russia

[2]Department of Statistical Modelling, St. Petersburg University, St. Petersburg, Russia

[3]Institute of Bioorganic Chemistry, Russian Academy of Sciences, Moscow, Russia

[4]Department of Computer Science and Engineering, University of California,
San Diego, USA

*These authors contributed equally to this work
**corresponding author, safonova.yana@gmail.com

# Supplemental Material A. Pitfalls of the clonal analysis of Rep-seq datasets without error correction

The standard Rep-seq protocols include an amplification step that introduces *amplification errors* that can be propagated by consecutive amplification cycles thus generating pseudo-diversity of an antibody repertoire (Pienaar et al. 2006; Bolotin et al. 2012). These errors are further compounded by *sequencing errors* in Illumina reads (estimated at 0.5% per base on average) leading to error-prone Rep-seq datasets, triggering errors in the constructed repertoires, and complicating the downstream analysis of antibodies. Amplification errors are often interpreted as erroneous branches in a clonal tree (Figure 1a), while sequencing errors make nearly all Rep-seq reads unique resulting in many false leaves in a clonal tree (Figure 1b). Thus, *error correction* of Rep-seq datasets with follow-up repertoire construction is a prerequisite for any downstream analysis of immunosequencing data.

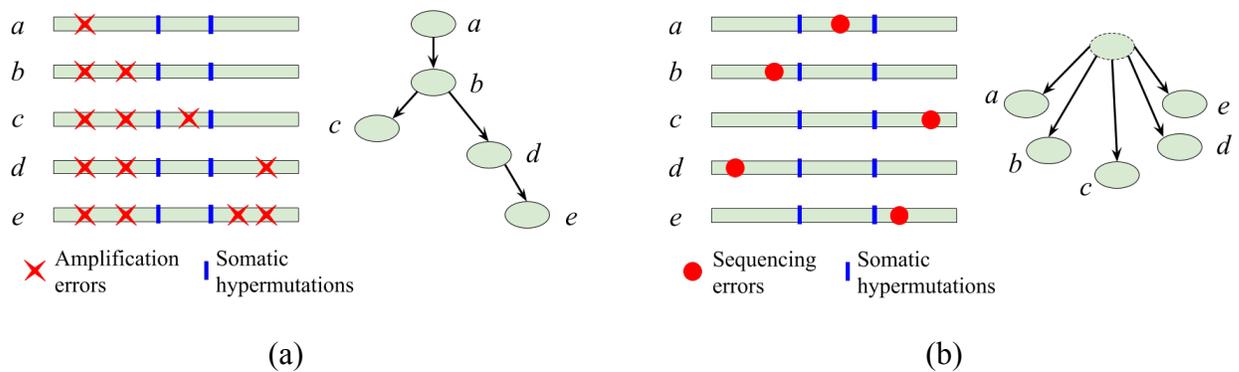

(a)                             (b)

**Figure 1: Pitfalls of the clonal analysis of Rep-seq datasets without error-correction.** In the absence of error correction, amplification (a) and sequencing (b) errors give rise to erroneous clonal trees (instead of classifying all five reads as a single antibody).

# Supplemental Material B. VJ Finder algorithm

VJ FINDER filters out contaminated and partial reads (i.e., reads that do not fully cover the V(D)J region) by aligning them against the Ig germline genes in the IMGT database (Lefranc et al. 2009). Alignment of immunosequencing reads against the germline database (known as *V(D)J labeling*) is a well-studied problem in immunoinformatics (Ye et al. 2013; Gaëta et al. 2007; Elhanati et al. 2015; Bonissone and Pevzner 2015). VJ FINDER addresses a more modest goal than V(D)J labeling as it attempts to label only V and J gene segments. Since VJ FINDER bypasses V(D)J labeling, it is much faster than existing V(D)J labeling tools (20X speed-up over IGBLAST), enabling analysis of large Rep-seq datasets. Our benchmarking of a high-throughput Rep-seq dataset demonstrated that VJ FINDER correctly aligned 99.9% of reads (see below).

Given a read and a germline gene, VJ FINDER recodes them as a string in the alphabet of their *k*-mers and computes the alignment between a read and a germline gene as the *longest common subsequence* between the resulting strings. For a given read, VJ FINDER selects *V* and *J hits* as a V and J segments with the best alignment score among all V and J germline genes, respectively. VJ FINDER also crops all reads by the first position of the V hit and the last position of the J hit.

**Computing alignment in VJ FINDER.** The VJ FINDER alignment algorithm consists of the following steps (Figure 2).

- FINDSHAREDKMERS step (Figure 3a) identifies positions of *k*-mers shared between *Read* and *Gene* and generates a set of *k-matches*, i.e., pairs (*r*, *g*), where *r* and *g* are starting positions of shared *k*-mers in *Read* and *Gene*, respectively.
- JOINCONSECUTIVEKMERS step (Figure 3b) joins consecutive *k*-matches into *blocks*. Each block is characterized by a triple (*r*, *g*, *l*), where *r* and *g* are starting positions of a shared *k*-mer in *Read* and *Gene*, respectively and *l* is the length of the block ($l \geq k$).

- CONSTRUCTCONSISTENCYGRAPH step (Figure 3c) constructs a directed acyclic *consistency graph* on blocks by connecting blocks $(r_1, g_1, l_1)$ and $(r_2, g_2, l_2)$ if $r_1 \leq r_2$ and $g_1 \leq g_2$. For each path in the consistency graph, we compute its score and length.

- FINDLONGESTPATH step finds a heaviest path in the consistency graph.

- CONSTRUCTALIGNMENT step (Figure 3d) constructs an alignment of *Read* and *Gene* by filling in the gaps between the blocks in the heaviest path. The alignment score is defined as the number of matches.

The sensitivity of the alignment algorithm is controlled by parameters $k_V$ (the *k*-mer size for V-alignment) and $k_J$ (the *k*-mer size for J-alignment). The default values of $k_V$ and $k_J$ are equal to 7 and 5, respectively.

**procedure VJ Finder**($Read, Gene, k$)

$kmers \leftarrow$ FindSharedKmers($Read, Gene, k$)

$blocks \leftarrow$ JoinConsecutiveKmers($kmers, k$)

$consGraph \leftarrow$ ConstructConsistencyGraph($blocks$)

$path \leftarrow$ FindLongestPath($consGraph$)

$alignment \leftarrow$ ConstructAlignment($path$)

**return** $alignment$

**Figure 2. VJ FINDER pseudocode.**

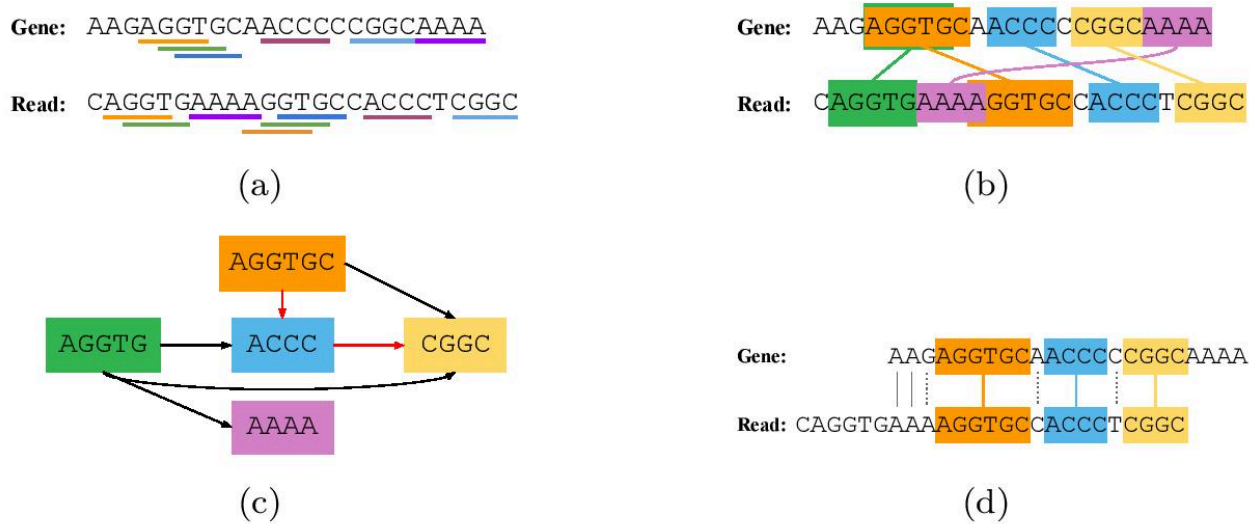

**Figure 3. Aligning *Read* against *Gene* using VJ FINDER.** (Upper left) Finding shared *k*-mers between *Read* and *Gene* and generating shared *k*-mers ($k = 4$). Each shared *k*-mer is represented as a pair $(r, g)$, where $r$ and $g$ are positions of this *k*-mer in *Read* and *Gene*, respectively. (Upper right) Joining consecutive *k*-mers into blocks. Each block is represented as a triple $(r, g, l)$, where $r$ and $g$ are positions in *Read* and *Gene* and $l$ is the length of the block ($l \geq k$). For example, since orange (AGGT), green (GGTG) and blue (GTGC) *k*-mers are consecutive in both the read and the gene, they are joined into a block AGGTGC of length 6. Note that blocks may overlap (e.g., a green block AGGTG is a prefix of an orange block AGGTGC in the read, but both of them have different positions on gene). (Lower left) Constructing consistency graph on blocks. Two blocks $(r_1, g_1, l_1)$ and $(r_2, g_2, l_2)$ are adjacent in the consistency graph if $r_1 < r_2$ and $g_1 < g_2$. For example, since this condition does not hold for orange (AGGTGC) and pink (AAAA) blocks, they are not connected by an edge in the consistency graph. The algorithm constructs a heaviest path in the consistency graph and computes its score as the total number of matching positions. A heaviest path with score 14 consists of blocks AGGTGC, ACCC and CGGC (corresponding edges are highlighted in red). (Lower right) Each path in the consistency graph corresponds to an alignment between the read and the gene with score defined as the number of matching positions (in our example, the alignment score is 16).

**Benchmarking VJ Finder.** We benchmarked VJ FINDER against IGBLAST on a highly mutated Rep-seq heavy chain dataset (referred to as LYMPH) from a cervical node of a patient suffering from multiple sclerosis (NCBI accession number SRR1383463). The dataset contains 1,852,661 paired-end reads, 76% of which were successfully merged into single full-length antibody sequences. We used IGBLAST as a reference V(D)J labeling tool for finding V and J hits for merged reads. VJ FINDER and IGBLAST took ∼1.5 hours and ∼ 30 hours, respectively to process this dataset.

A read is classified as a *contaminant* if *e*-value of its alignment (as computed by IGBLAST) exceeds 0.001. VJ FINDER found all 37,435 contaminant reads and 28,537 partial reads, i.e., reads that contain only partial V or J genes. For the remaining reads, we compared V and J hits as well as their starting and ending positions reported by VJ FINDER and IGBLAST. VJ FINDER results were inconsistent with IGBLAST output only for 0.1% of reads.

# Supplemental Material C. Constructing the Hamming graph

Figure 4 presents the pseudocode of the Hamming graph construction algorithm while Figure 5 presents an example of a Hamming graph construction.

    **procedure HGConstructor**($Reads, k, \tau$)

        $KmerIndex \leftarrow \emptyset$

        **for each** $Read$ in $Reads$ **do**

          **for each** $k$-mer $\kappa$ in GenerateKmers($Read$) **do**

            $KmerIndex(\kappa) \leftarrow Read$

        $Graph \leftarrow \emptyset$

        **for each** $Read_1$ in $Reads$ **do**

          $Kmers_{Read_1} \leftarrow$ GenerateKmers($Read_1$)

          $Minimizers_{Read_1} \leftarrow$ Minimizers($Kmers_{Read_1}$)

          **for each** $\mu$ in $Minimizers_{Read_1}$ **do**

            **for each** $Read_2$ **in** $KmerIndex(\mu)$ **do**

              **if** $d(Read_1, Read_2) \leq \tau$ **then**

                add edge $(Read_1, Read_2)$ to $Graph$

        **return** $Graph$

**Figure 4. Pseudocode of the Hamming graph construction algorithm.** The GENERATEKMERS function generates all unique *k*-mers in a read and is used for generating KMERINDEX (a map from *k*-mers to indices of reads containing them) and computing *k*-mer multiplicities (the number of reads containing a *k*-mer). HGConstructor calls the Minimizers algorithm as a filtration procedure (see sub-section "The Minimizers algorithm" below).

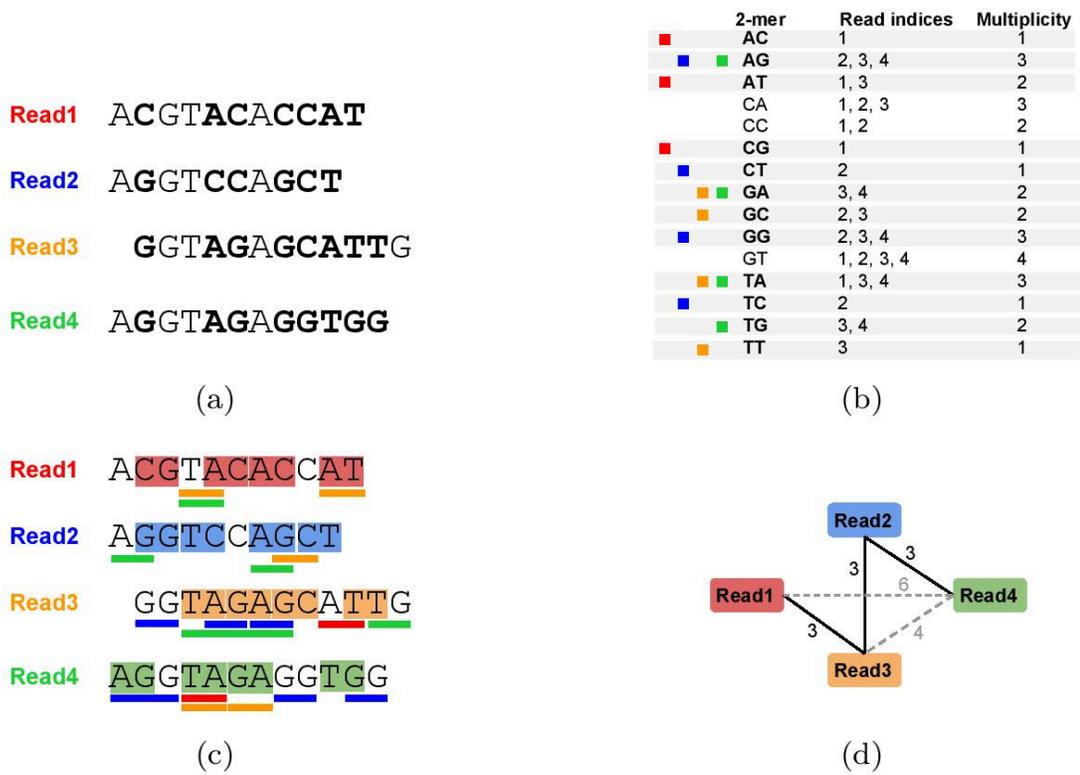

**Figure 5. Constructing the Hamming graph using the minimizers strategy ($\tau = 3$, $k = 2$).** (Upper left) Multiple alignment of reads. Columns corresponding to mismatches are highlighted in bold. (Upper right) The algorithm computes minimizers using a *k*-mer index that stores indexes of reads sharing each *k*-mer occurring in reads. For example, AC is presented in $Read_1$ only, while GT is presented in all 4 reads. Multiplicity of a *k*-mer is defined as the number of reads containing it. The algorithm finds minimizers in each read, i.e., $\tau + 1$ non-overlapping *k*-mers with minimal total multiplicity. For example, GG, TC, AG, and CT (represented by blue squares in the *k*-mer index) represent minimizers for the blue $Read_2$ with total multiplicity 8. (Lower left) To compute pairs of reads connected by edges in the Hamming graph, the algorithm finds $\tau + 1$ non-overlapping minimizers for each read (these minimizers are colored by the colors of the reads they occur in). For example, CG, AC, AC, and AT are minimizers for the red $Read_1$ (note that AC appears twice in the list of minimizers). Minimizers occurring in other reads are shown by colored lines with colors inherited from the read that gave rise to this minimizer. For example, the blue $Read_2$ contains minimizers of the orange $Read_3$ (GC) and the green $Read_4$ (AG), but does not contain minimizers of the red $Read_1$. The algorithm connects two reads (vertices in the Hamming graph) by an edge if they share a minimizer. For example, there is no edge between $Read_1$ and $Read_2$. (Lower right) The resulting Hamming graph consists of three edges: ($Read_1, Read_3$),

($Read_2, Read_3$), and ($Read_2, Read_4$) (all these edges have weight 3). Grey dashed edges correspond to pairs of reads for which the computed Hamming distance exceeds the threshold exceeds $\tau$.

**The Minimizers algorithm.** Figure 6 presents the pseudocode of the recursive MINIMIZERS algorithm. Note that since MINIMIZERS works with prefixes in the $k$-mer list (READKMERS), it can also be implemented using dynamic programming with running time and memory footprint $\mathbf{O(|ReadKmers| \cdot (\tau + 1))}$. The minimizers approach can be generalized for the edit distance to account for insertions and deletions introduced at the amplification stage.

> **global** $k, KmerIndex$
> **procedure Minimizers**($ReadKmers, \tau$)
>     **if** $\tau + 1 = 0$ **then**
>         **return** $\emptyset$
>     $n \leftarrow |ReadKmers|$
>     **if** $n = \tau \cdot k + 1$ **then**
>         **return** $ReadKmers_1, ReadKmers_{k+1}, \dots, ReadKmers_{\tau \cdot k + 1}$
>     $Kmers_{skip} \leftarrow \text{Minimizers}(ReadKmers[1:n-1], \tau + 1)$
>     $Kmers_{take} \leftarrow \text{Minimizers}(ReadKmers[1:n-k], \tau) \cup \{ReadKmers_n\}$
>     **if** $\text{SumMult}(Kmers_{skip}, KmerIndex) < \text{SumMult}(Kmers_{take}, KmerIndex)$ **then**
>         **return** $Kmers_{skip}$
>     **Else**
>         **return** $Kmers_{take}$

**Figure 6. Pseudocode of the Minimizers algorithm.** For all $k$-mers extracted from a read, the algorithm selects $\tau + 1$ $k$-mers with minimal total multiplicity (multiplicity of a $k$-mer is stored in KMERINDEX). The SUMMULT procedure computes the total multiplicity for a set of $k$-mers.

**Benchmarking strategies.** We evaluated the minimizers strategy, the Knuth filtering strategy (Knuth 1998), its modification implemented in IGREPERTOIRECONSTRUCTOR (Safonova et al. 2015), and the brute-force algorithm based on the average number of Hamming distance computations per read.

The Knuth strategy works only for computing the standard Hamming distance, i.e., for sequences starting from the same positions. In the case of known germline genes, the Knuth strategy can be applied since we crop reads by the first position of alignment against the V segment. However, if germline segments are not known or if the V alignment is ambiguous (e.g., for highly hypermutated antibodies), the starting position of an antibody sequence is difficult to define. In such cases, the *extended Hamming distance* (Safonova et al. 2015) should be applied instead of the standard Hamming distance.

**Benchmarking datasets.** Benchmarking was performed on SIMULATED, SYNTHETIC, REAL, and LYMPH datasets.

We also generated REAL GERMLINE and REAL FIXED datasets based on the REAL dataset. The REAL GERMLINE dataset models a situation when sequencing of antibody molecules starts from the middle of a V segment and is obtained by replacing the first 150 nt of each read with the 150-nt prefix of the corresponding V hit. The REAL FIXED dataset models a large group of naïve antibodies sharing a V segment and is obtained by replacing the first 150 nt of each read with a fixed 150 nt-sequence.

**Benchmarking results.** Table 1 and Figure 7 present benchmarking results and illustrate that on average the minimizer strategy is roughly 30 times faster than the standard Knuth strategy (167X speed-up for the REAL FIXED dataset) and roughly 80 times faster than the brute-force approach (106X speed-up on the REAL dataset).

Figure 8 (a) shows the number of the Hamming distance computations for $\tau = 1, 2, 3, 4$ and for different $k$-mers sizes (REAL dataset). For $\tau = 4$, the minimum number of the Hamming distance computations is achieved with $k = 10$ while for $\tau < 4$ the best results are achieved with longer $k$-mers, e.g., 48-mers for $\tau = 1$. Figure 8 (b) shows positions of minimizers (for $k = 10$ and $\tau = 4$) and illustrates that the minimizers strategy selects $k$-mers mostly from CDRs, the most diverged regions of immunoglobulins.

|  |  | Average # HD computations per read | | | |
|---|---|---|---|---|---|
| Dataset | # reads | Minimizers ($k = 10$) | Knuth ($HD$) | Knuth ($\widetilde{HD}$) | Brute-force |
| REAL | 264,941 | **1246** | 3274 | 23,215,090 | 132,470 |
| SIMULATED, #errors = 1 | 34,398 | **640** | 647 | 6,882,160 | 17,199 |
| SYNTHETIC, #errors = 1 | 598,669 | **3640** | 3856 | 196,602,478 | 299,334 |
| LYMPH | 1,331,032 | **4472** | 5624 | 10,352,500 | 665,516 |
| REAL GERMLINE | 152,361 | **1014** | 9319 | 41,137,701 | 76,180 |
| REAL FIXED | 150,820 | **1005** | 168,20 | 634,800,582 | 75,410 |

**Table 1. Benchmarking of various Hamming graph construction strategies.** The average number of the Hamming distance computations per read for the minimizer strategy (the "Minimizers" column), the Knuth filtration strategy (the "Knuth ($HD$)" column), modification of the Knuth strategy for the extended Hamming distance (the "Knuth ($\widetilde{HD}$)" column), and the brute-force algorithm (the "Brute-force" column). The best values in each row are highlighted in bold. The distance threshold $\tau$ is equal to 4. Identical Rep-seq reads were glued before benchmarking to reduce the number of the Hamming distance computations.

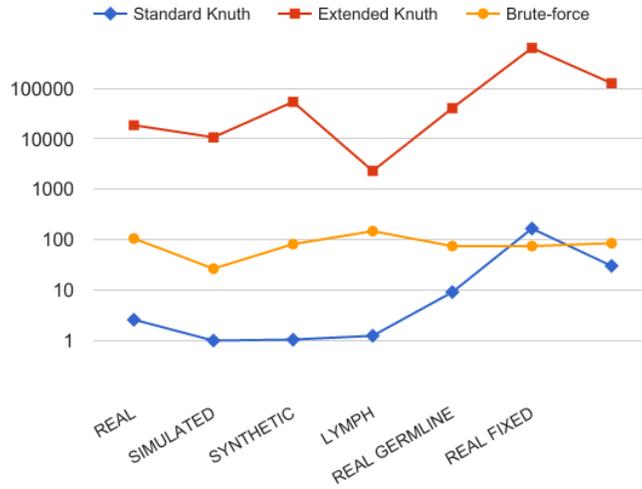

**Figure 7. Speed-up of the minimizers strategy versus the standard Knuth filtration strategy (blue diamond), modification of the Knuth strategy for computing the extended Hamming distance (red square), and the brute-force strategy (orange circle).** The speed-up values were computed based on the average number of the Hamming distance computations (Table 1. ). The *y*-axis is represented in the logarithmic scale.

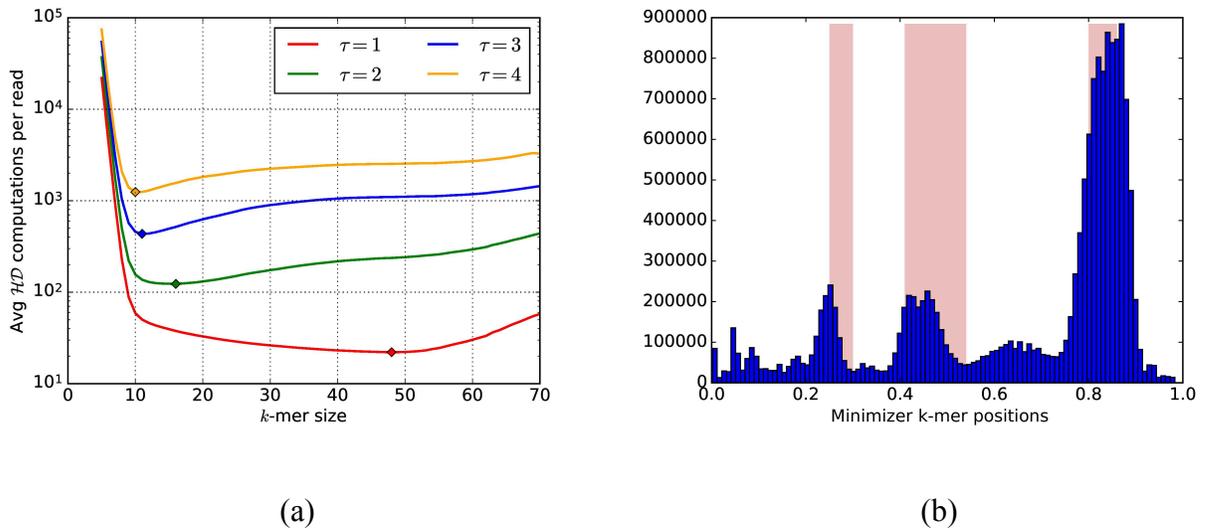

(a) (b)

**Figure 8. Applying the minimizers strategy to the REAL dataset.** (a) The average number of the Hamming distance computations per read for $\tau = 1, 2, 3$, and $4$. For each $\tau$, a point corresponding to the minimal value is marked by a diamond. Optimal values of $k$ for $\tau = 1, 2, 3$, and $4$ are 48, 16, 11, and 10, respectively. (b) The distribution of the relative positions of minimizers for $k = 10$ and $\tau = 4$. Red bars correspond to the relative positions of CDRs.

**Applying the minimizers algorithm to Alu repeats.** About 11% of the human genome consists of *Alu repeats*, the most common repeat in primate genomes. Alu repeats evolved according to a *clonal tree*, i.e., a tree where the known Alu repeats represent both leaves and internal nodes (Price et al. 2004).

We addressed the problem of Alu subfamily classification by constructing the Hamming graph for three Alu subfamilies: *AluJr4* (8863 sequences), *AluSc5* (5201 sequences) and *AluYm1* (4335 sequences). For each subfamily we computed the value of *k* minimizing the number of the Hamming distance computations for $\tau = 10$ (30-mers, 35-mers, and 35-mers for *AluJr4*, *AluSc5*, and *AluYm1*, respectively). Figure 8a-c show the distributions of positions of minimizers for each of three analyzed subfamilies and reveal surprising features of each subfamily with respect to the distribution of mutations, e.g., the prefix of *AluJr4* is much more conserved than the prefix of *AluSc5* and *AluYm1*.

We also mixed sequences from all subfamilies and constructed the Hamming graph using $\tau = 20$. As Figure 8d illustrates, each of the four non-trivial connected components in the Hamming graph contains Alu repeats from the same subfamily. The split of the *AluSc5* subfamily into two non-trivial connected components is likely explained by indels in alignment of sequences from the *AluSc5*.

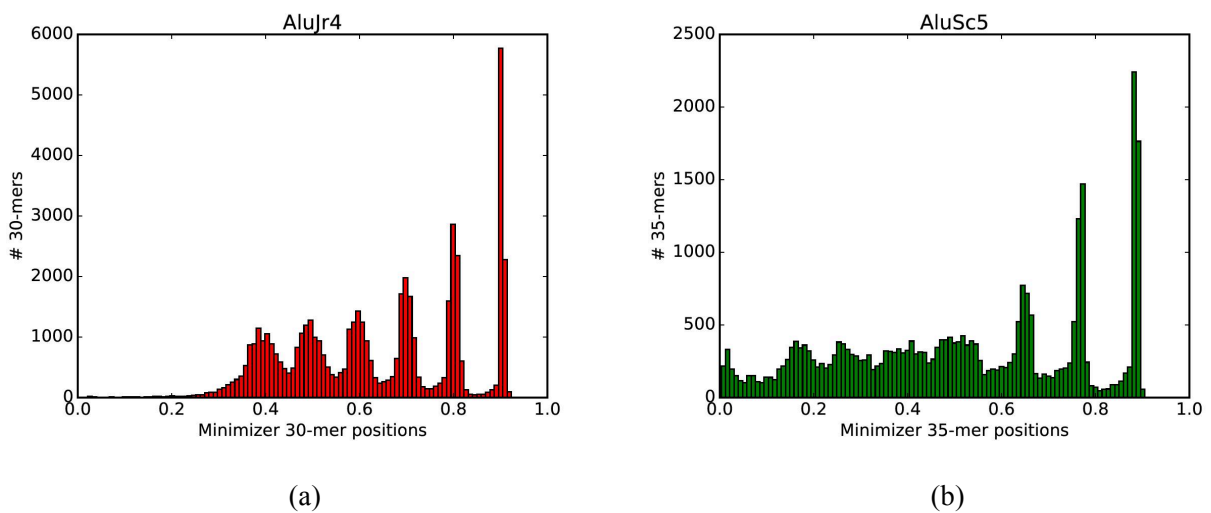

(a)          (b)

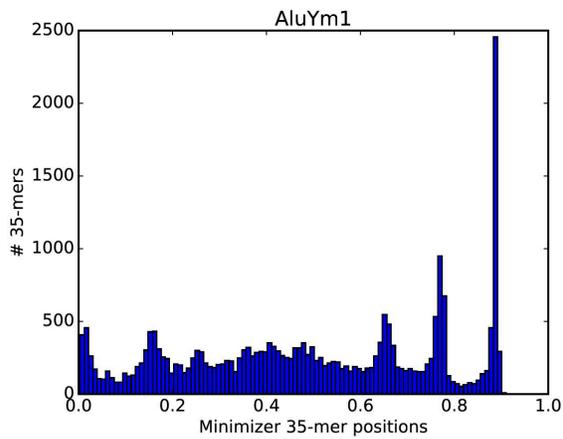
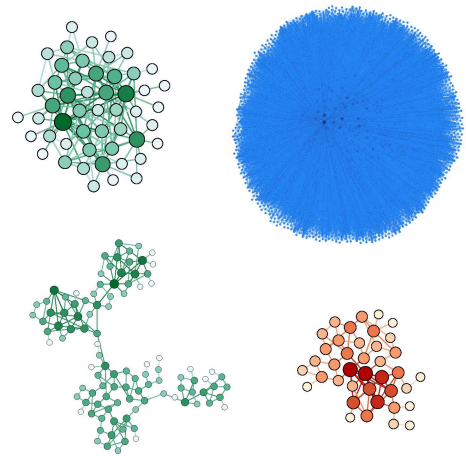

(c)                               (d)

**Figure 8. Applying the minimizers strategy to AluJr4 (red), AluSc5 (green), and AluYm1 (blue) subfamilies.** (a), (b), and (c) show distributions of positions of minimizers for *AluJr4*, *AluSc5*, and *AluYm1*, respectively, while (d) shows the Hamming graph with four non-trivial connected components constructed for the mixture of all three subfamilies (with $\tau = 20$). Each connected component consists of sequences from the same subfamily. *AluSc5* subfamily corresponds to two connected components.

# Supplemental Material D. Selecting BARCODEDIGREC parameters

$\tau_{cluster}$ parameter controls read clustering for each barcode as specified in the "Grouping by barcodes and resolving collisions" step of BARCODEDIGREC. Two reads are placed in the same cluster if the edit distance between them does not exceed $\tau_{cluster}$. To select an optimal value of $\tau_{cluster}$, we analyzed the REAL dataset and computed distributions $D_1$ (distances between reads that belong to identical antibodies) and $D_2$ (distances between randomly chosen reads). To compute the $D_1$ distribution, we calculated edit distances between all pairs of reads sharing a barcode. To compute the $D_2$ distribution, we calculated edit distances for all pairs of clusters from a repertoire reported by BARCODEDIGREC. Figure 9 presents a joint histogram of $D_1$ (red) and $D_2$ (blue) distributions and illustrates that $\tau_{cluster} = 20$ separates receptor sequences that belong to the same antibody clusters from randomly selected receptor sequences.

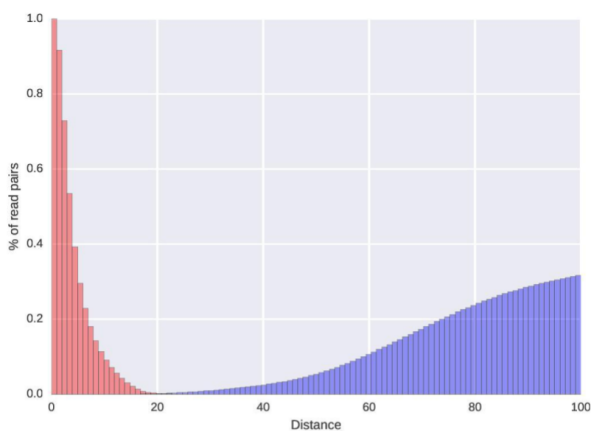

**Figure 9. Distributions of edits distances between Rep-seq reads corresponding to identical antibodies (red) and randomly selected receptor sequences (blue).** The red histogram was computed based on edit distances for all pairs of reads sharing a barcode. Although some of these pairs may correspond to distinct antibodies due to barcode collisions, their contribution to this histogram is expected to be small. The blue histogram was computed based on the edit distances between all pairs of cluster sequences as defined by BARCODEDIGREC with aggressive clustering parameters for the REAL dataset.

$\tau_{umi}$ parameter is used for correcting errors within barcodes as specified in the "Detecting and resolving barcode errors" step of BARCODEDIGREC. BARCODEDIGREC glues clusters that belong to similar barcodes (differing in at most $\tau_{umi}$ positions) if the edit distance between them does not exceed $\tau_{cluster}$. Increasing the $\tau_{umi}$ parameter results in correcting more barcode errors but may erroneously glue different but clonally related receptor sequences. Thus, the larger $\tau_{umi}$, the higher the risk to glue clonally related clusters from similar barcodes. To avoid such effects, we select a small default value $\tau_{umi} = 1$. Increasing $\tau_{umi}$ from 1 to 2 leads to gluing ~1000 clusters in the REAL dataset. Although gluing these clusters does not significantly corrupt the diversity of the REAL dataset (since most of them present singletons), increasing $\tau_{umi}$ may be detrimental for highly mutated repertoires.

$\tau_{igrec}$ parameter defines the value of the $\tau$ parameter used by IGREC for the Hamming graph construction at the "Merging similar clusters" step of BARCODEDIGREC. We selected a small value of this parameter to avoid construction of edges between similar sequences thus preserving clonal diversity. Although the consensus sequences constructed at the previous steps of the BARCODEDIGREC algorithm are rather accurate, some of them may contain a single mismatch that we would like to correct. These incorrect sequences will be grouped into a correct cluster corresponding to a star-like component in the Hamming graph with $\tau = 1$. However, since we want IGREC to glue these components into a single cluster in a final repertoire and to turn them into dense subgraphs, we set the default value of $\tau_{igrec}$ to 2.

# Supplemental Material E. Abundance analysis with IGQUAST

It turned out that the existing repertoire construction tools often result in *overestimation* (abundance of the constructed cluster exceeds the abundance of the reference cluster) and sometimes results in *underestimation* (abundance of the constructed cluster is smaller than the abundance of the reference cluster). Overestimation is a result of *overcorrection* (gluing several reference clusters into a single cluster in the constructed repertoire), while underestimation is caused by *undercorrection*. Analysis of cluster abundances with IGQUAST revealed that most abundances are underestimated (on both barcoded and non-barcoded datasets). This effect can be explained by a tendency of repertoire construction tools to split a reference cluster into a large cluster with the correct consensus and multiple small erroneous clusters.

**Defining matches between the constructed and the reference repertoires.** To define matches between the constructed and the reference repertoires, we take into account that sequences in these repertoires often have slightly different lengths because various repertoire construction approaches crop cluster sequences differently (e.g., by the end of CDR3 or by the end or the *framework segment 4* (FR4)). While the starting positions of a cluster sequence coincides with the starting position of the corresponding V gene segment, the ending positions of a cluster sequence may differ from the ending position of the corresponding V gene segment. Thus, we define the match between the constructed and reference sequences as the exact match between the shorter sequence and the prefix of the longer sequence of the same length.

**Abundance analysis on non-barcoded datasets.** Figure 10 presents the results of three repertoire construction tools on the non-barcoded version of the REAL dataset (i.e., without using information about barcodes). The median of the ratios of the constructed abundances to the reference abundances varied from 0.9 for IGREC (Figure 10a), to 0.8 for MIXCR (Figure 10b), and to 0.47 PRESTO (Figure 10c). Both IGREC and MIXCR generate few overcorrected clusters (points above $y = x$ line). For each overcorrected cluster, we computed the *overcorrection coefficient* as the ratio of the constructed abundance to the reference abundance. For IGREC and MIXCR, the average values of the overcorrection coefficient are 1.25 and 2.1, respectively. Since PRESTO does not correct errors in reads, it underestimates abundances of all clusters.

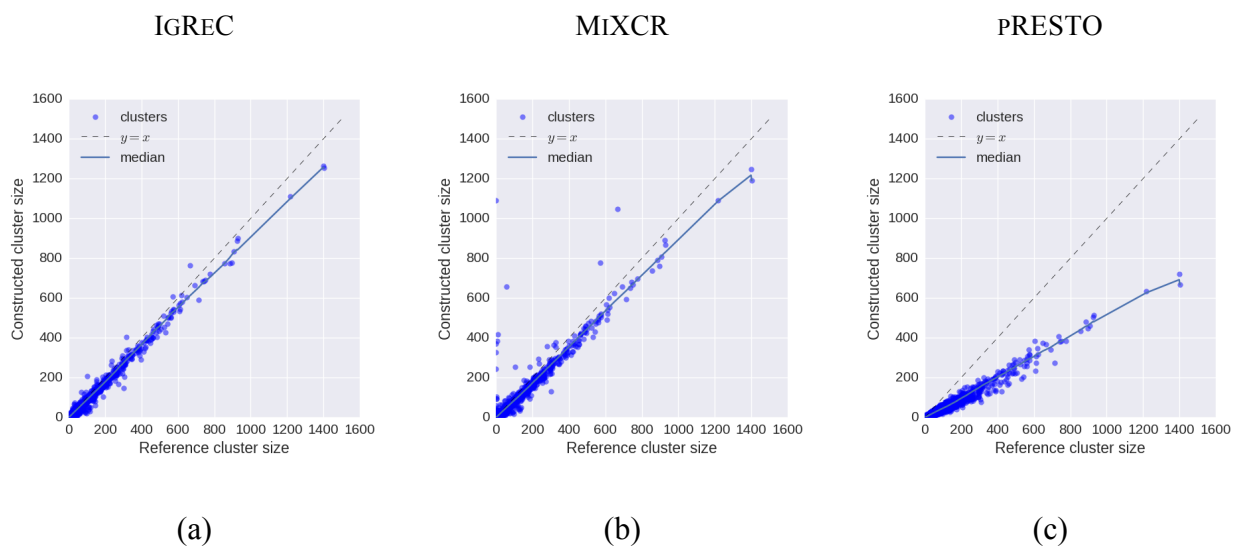

(a)          (b)          (c)

**Figure 10. Abundance plots for the repertoires constructed for the REAL dataset by IGREC (a), MIXCR (b), and PRESTO (c).** The plots show dependency between abundances in the reference (*x*-coordinate) and the constructed (*y*-coordinate) repertoires. Each point corresponds to a cluster with the correctly reconstructed sequence. Most abundances of the constructed clusters are underestimated, i.e., points lie below the $y = x$ line. For each cluster, we computed the ratio of the constructed abundance to the reference abundance and computed a *cumulative median* as the array of medians for suffixes of sorted ratio values shown as blue curves. The curves are close to linear for all tools. There are several points (especially, for MiXCR) above the $y = x$ line corresponding to overcorrected clusters. Median ratios for large clusters (with abundances at least 5) are 0.9, 0.8, and 0.47 for IGREC, MIXCR, and PRESTO, respectively.

**Abundance analysis on barcoded datasets.** Figure 11 presents abundance analysis for the SIMULATED BARCODED dataset (2 errors per read) and illustrates that BARCODEDIGREC and MIGEC + MIXCR show the best results in terms of evaluating abundances. PRESTO and IgReC in blind mode result in undercorrected repertoires. All tools do not overcorrect clusters.

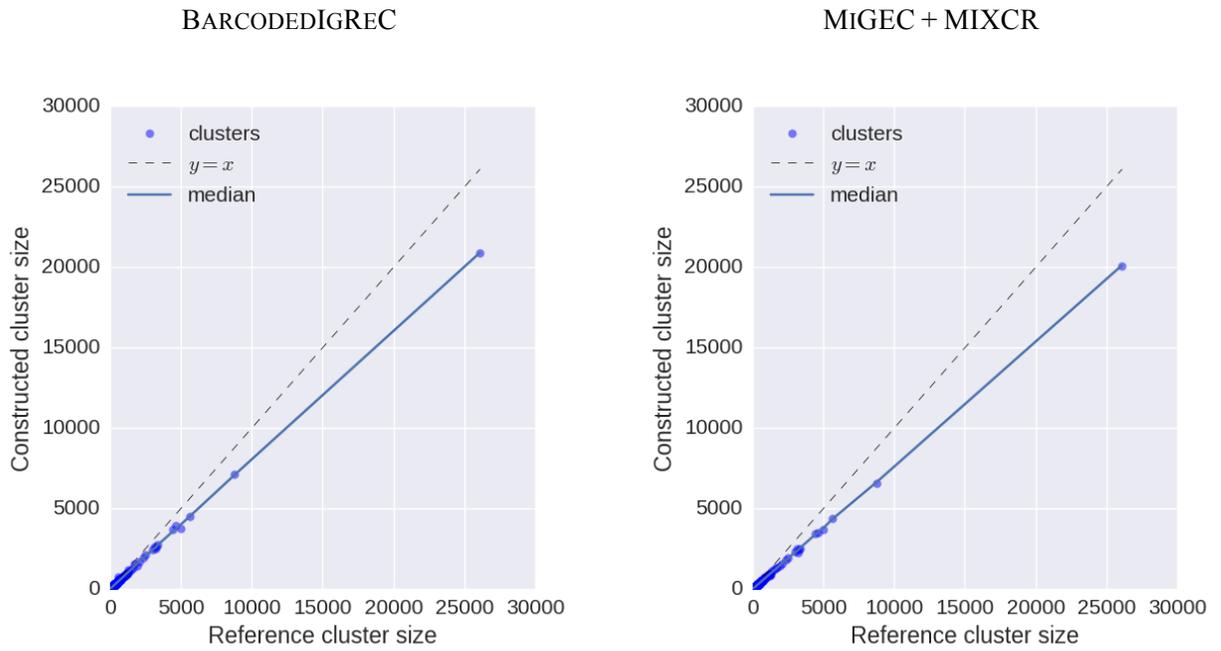

(a) BARCODEDIGREC

(b) MIGEC + MIXCR

PRESTO

IGREC (in blind mode)

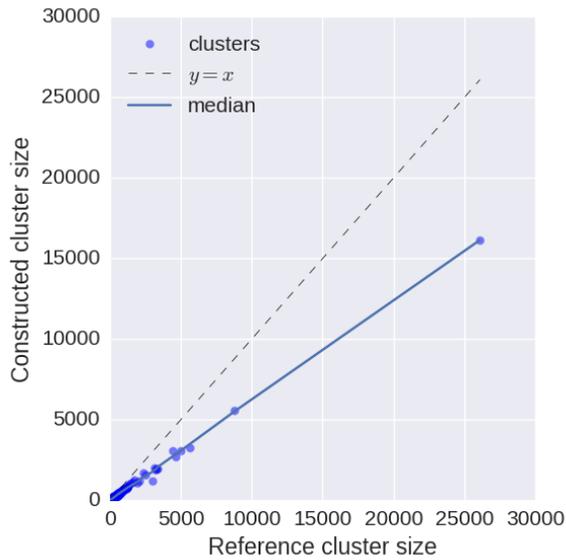 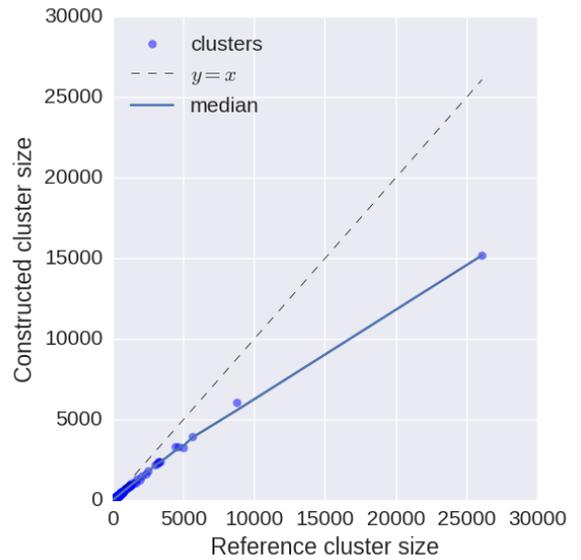

(c)                                         (d)

**Figure 11. Abundance plots for repertoires constructed for the SIMULATED BARCODED dataset (2 errors per read) by BARCODEDIGREC (a), MIGEC + MIXCR (b), PRESTO (c), and IGREC (d) tools.** The dependence between abundances in the reference and the constructed repertoires is close to linear for all tools. Ratios of the constructed abundances to the reference abundances have small variance, and there are no overcorrected clusters. Median ratios for large clusters (of abundance at least 5) are 0.81, 0.83, 0.58, and 0.56 for BARCODEDIGREC, MIGEC + MIXCR, PRESTO, and IGREC, respectively.

**Abundance analysis on barcoded datasets using barcode counting.** For each cluster in a constructed repertoire we can compute its *barcode abundance* (the number of RNA molecules contributing to this cluster) and use the barcode abundances as references to evaluate the standard abundances (the number of reads in the cluster). Figure 12a compares barcode abundance and read abundance for the REAL repertoire computed by BARCODEDIGREC and reveals that although a large barcode abundance typically corresponds to a large read abundance, the variance of read abundances for a given barcode abundance is very high. For example, for the barcode abundance 10, the read abundances vary from 10 to 1000. Thus, barcode abundances represent the method of choice in immunogenomics studies aimed at accurate quantitation of RNA molecules.

We also analysed barcode abundance reconstructed by BARCODEDIGREC using abundance plots where the standard abundance was replaced by the barcode abundance. Figure 12b shows the barcode abundance plot for the SIMULATED BARCODED dataset (2 errors per read). The high value of median ratio (0.9 for the barcode abundance versus 0.81 for the read abundance) confirms that BARCODEDIGREC accurately reconstructs repertoires in terms of both sensitivity/precision and barcode abundance.

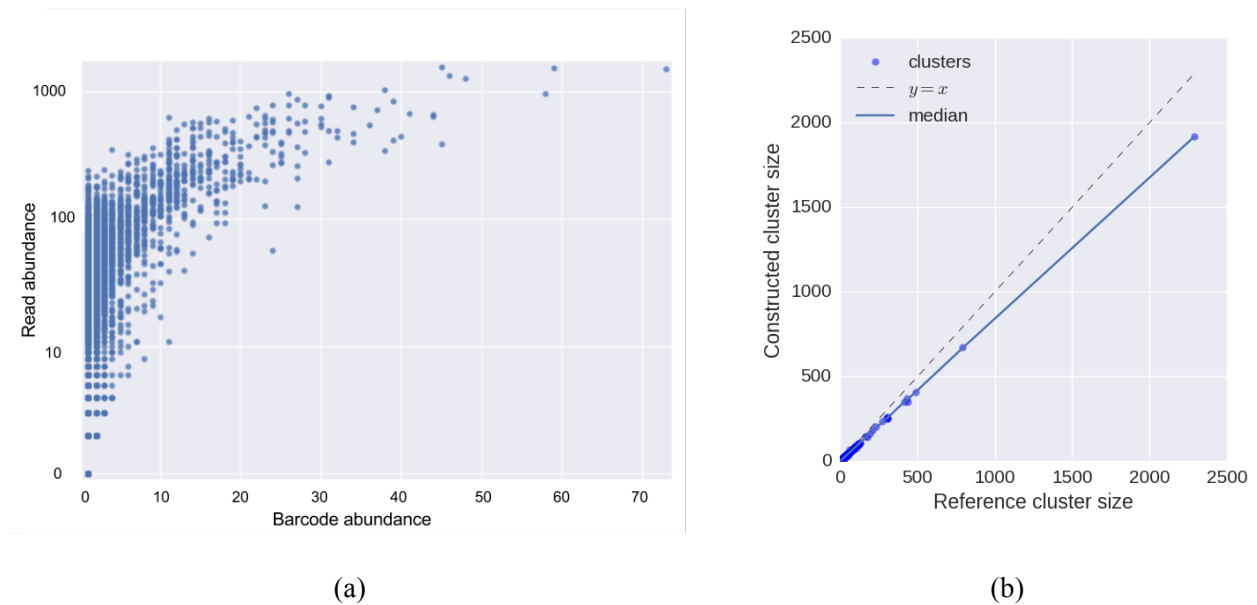

(a)                                             (b)

**Figure 12. Barcode abundance analysis of repertoires constructed from barcoded Rep-seq datasets.** (a) Barcode abundance versus read abundance for clusters in the REAL repertoire computed by BARCODEDIGREC. (b) The barcode

abundance plot for the SIMULATED BARCODED repertoire (2 errors per read) constructed by BARCODEDIGREC. Median ratio for large clusters is 0.9.

# Supplemental Material F. Detecting overcorrection with IGQUAST

**Reference-based detection of overcorrection.** Figure 13a illustrates that the SYNTHETIC repertoire corresponding to a strong immune response contains many overcorrected clusters (53% of all large clusters), while the REAL repertoire corresponding to a healthy individual contains very few overcorrected clusters (5%). We analyzed overcorrected clusters using information about the partition of the reference reads and the partition of reads constructed by IGREC.

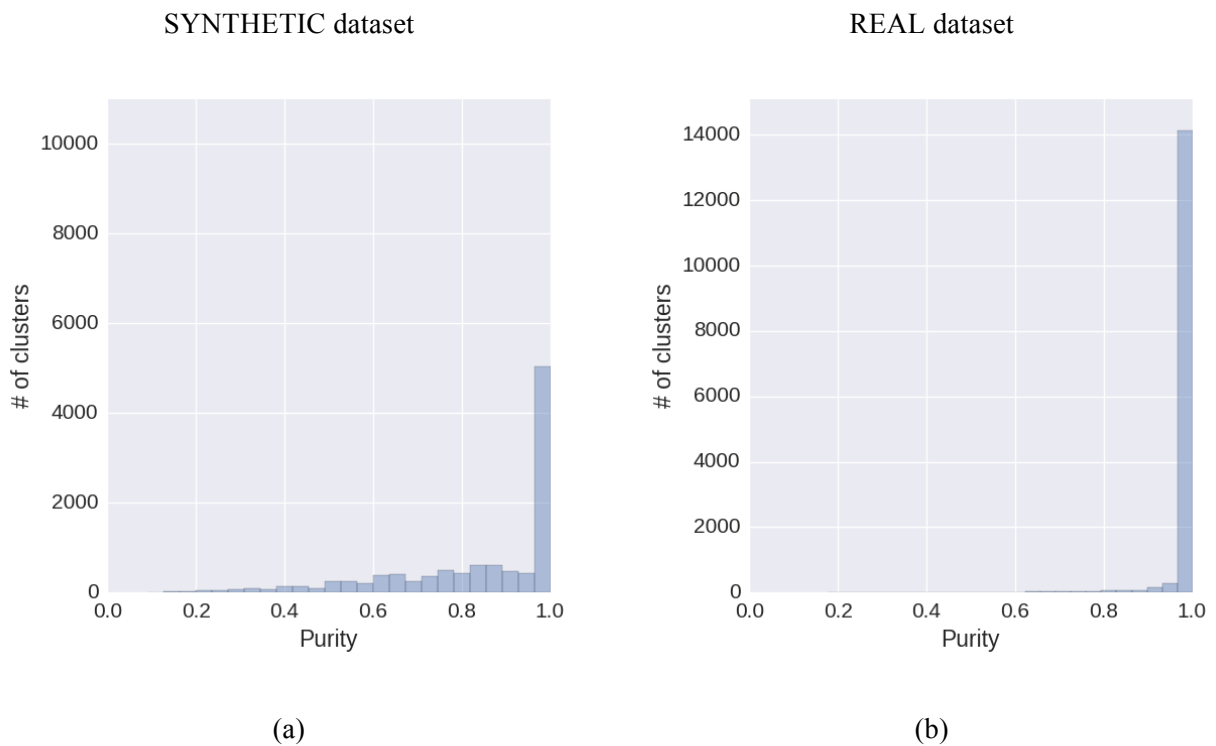

(a)            (b)

**Figure 13. Distribution of purity using reference-based analysis of large clusters for the SYNTHETIC (a) and the REAL (b) repertoires.** Large clusters are defined as clusters with at least 5 reads.

**Reference-free detection of overcorrection.** Analysis of overcorrected clusters can be extended to the case when the reference repertoire is unknown. IGQUAST aligns all reads from each large cluster against the consensus sequence of the cluster and analyzes all columns in the resulting multiple alignment. For each column, it computes the *discordance*, the fraction of reads corresponding to the second most abundant nucleotide in this column. The *discordance* of a cluster in the constructed repertoire is defined as the maximal value of discordances among all columns. Gluing multiple similar antibodies into a single cluster results in high discordance and reveals overcorrection (Figure 14). IgQUAST reports clusters with large discordance as potentially overcorrected.

We analyzed overcorrected clusters constructed by IGREC for the SYNTHETIC and REAL datasets in the reference-free mode. For each constructed cluster with the second vote exceeding 5, we computed whether splitting this clusters by the corresponding column changes the sensitivity. For the SYNTHETIC repertoire, splitting 8705 / 1767 / 522 clusters results in unchanged/increased/decreased sensitivity of the repertoire (Figure 15a). Thus, it makes sense to split clusters with large second vote for highly mutated repertoires (e.g., repertoires of vaccinated or infected individuals) since it allows one to reconstruct more clusters and thus improve follow-up clonal analysis. For the REAL repertoire, splitting 13,588 / 339 / 1165 clusters results in unchanged/increased/decreased sensitivity of the repertoire (Figure 15b), implying that it does not make sense to split clusters with large second vote in this case.

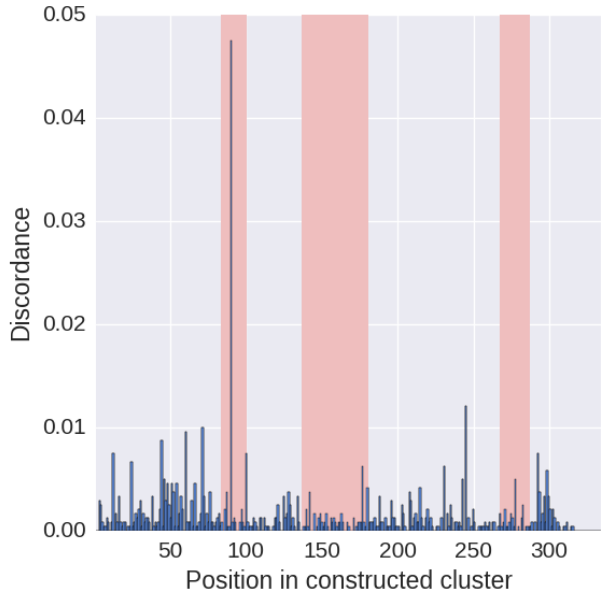

**Figure 14: Reference-free detection of overcorrected clusters**. Plot showing the number of reads with the given position (*x*-coordinate) maximizing the discordance. Red bars correspond to positions of CDRs. The peak at position 101 represents ~5% of reads in the constructed clusters and likely corresponds to overcorrection. Data is shown for the repertoire constructed by the IgReC tool for the REAL dataset.

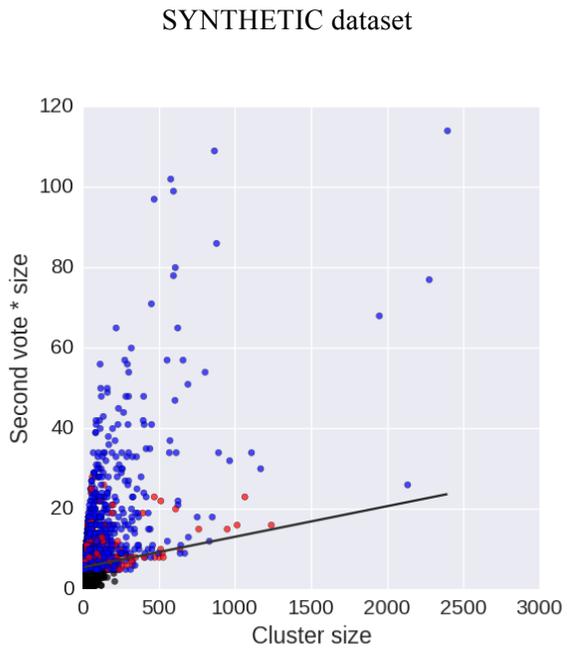
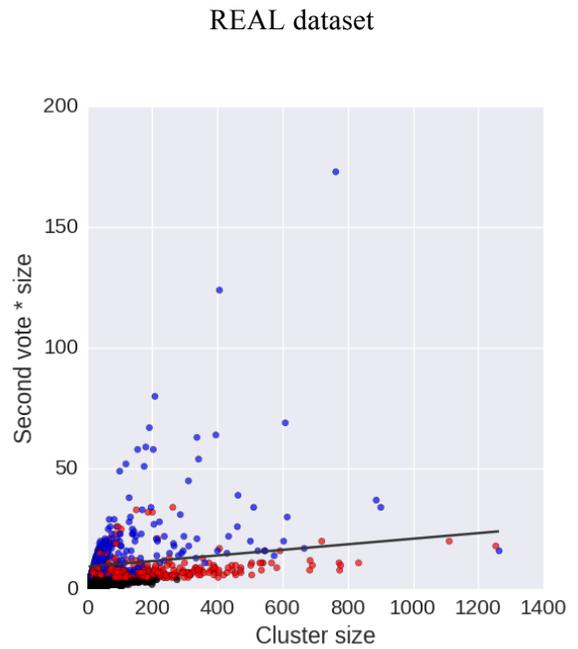

(a)                                                   (b)

**Figure 15. Plots showing dependence of second vote on the cluster size for the SYNTHETIC (a) and the REAL (b) repertoires constructed by IGREC.** Each point represents a cluster. A point is colored blue/red if splitting the corresponding cluster by a column with the second vote increases/decreases the sensitivity of repertoires. Splitting black clusters does not change the sensitivity of the repertoire.

# Supplemental Material G. IGDIVERSITYANALYSER

V(D)J recombination generates the primary diversity of adaptive immune repertoire, while the secondary diversification is achieved by clonal expansion and somatic hypermutations. Diversity analysis of a repertoire can be used for evaluating the state of adaptive immune system, e.g., an antibody repertoire of an infected/vaccinated patient may contain large clonal families (that are reactive to an attacking antigen), while a repertoire of a healthy individual peripheral blood typically consists of small clonal families resulting from recent immune responses or presenting a memory pool. Analysis of CDRs represents another aspect of diversity analysis (Murugan et al. 2012) (Elhanati et al. 2015) (Lindner et al. 2012).

Many tools for analyzing diversity focus on T cell repertoires, e.g., TCR (Nazarov et al. 2015) and TCRKLASS (Yang et al. 2015). Since these tools represent TCR repertoire as a set of their CDR3 sequences, they are not applicable to the full-length antibody repertoires. Popular IMGT/HIGHV-QUEST (Li et al. 2013) tool analyzes diversity of antibody repertoires, but does not have publicly available stand-alone versions and lack ability to analyze large Rep-seq datasets. In contrast, MIXCR (Bolotin et al. 2015) and ALAKAZAM (Gupta et al. 2015) are publicly available tools for diversity analysis of full-length antibody repertoires. VDJTOOLS (Shugay et al. 2015) performs post-processing of MIXCR output and reports various diversity statistics.

IGDIVERSITYANALYZER computes various statistics of a repertoire and presents them as an HTML report. IgDIVERSITYANALYZER is launched at the last stage of IgReC and is also available as a stand-alone tool. Figure 16 presents IGDIVERSITYANALYZER output for a repertoire constructed by IGREC for a Rep-seq library from an HIV-infected patient (94-th week of disease) described in Schanz et al. 2014.

IgDIVERSITYANALYZER launches VJ FINDER to generate VJ alignments of all reads and uses them for CDR and SHM annotations. It relies on canonical CDR annotations of V and J germline segments (as reported by IGBLAST) and uses the constructed alignments to projects positions of the germline CDRs into receptor sequences. By default, IGDIVERSITYANALYZER uses the IMGT notation (Lefranc et al. 2009) for CDR labeling, but also provides an option of using the KABAT notation (Wu and Kabat 1970). It further classifies alignment columns with differing receptor and germline sequences as SHMs. IGDIVERSITYANALYZER is unable to detect SHMs that change an already mutated base into the germline nucleotide, since detection of such SHMs requires advanced clonal analysis.

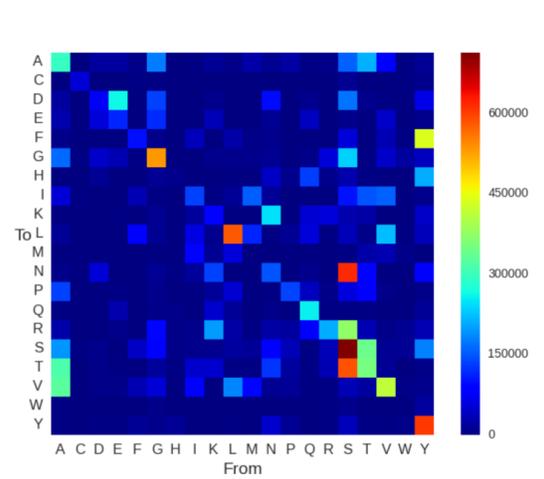 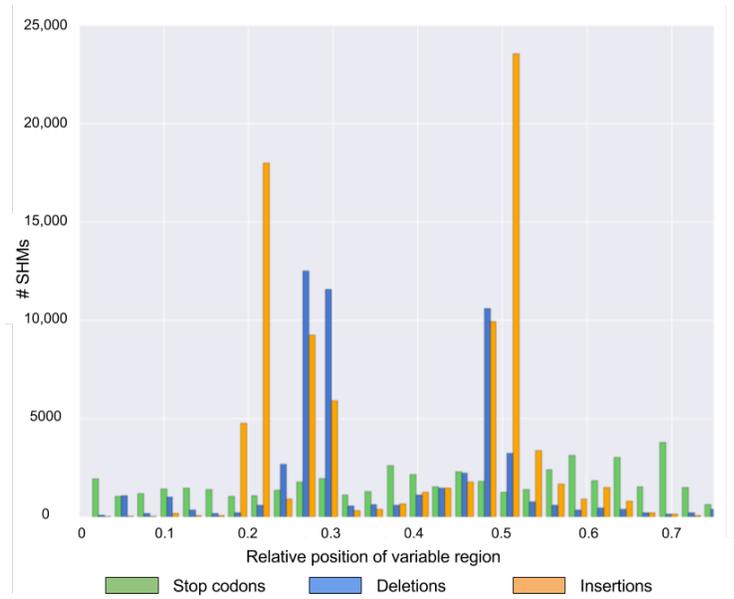

(a) (b)

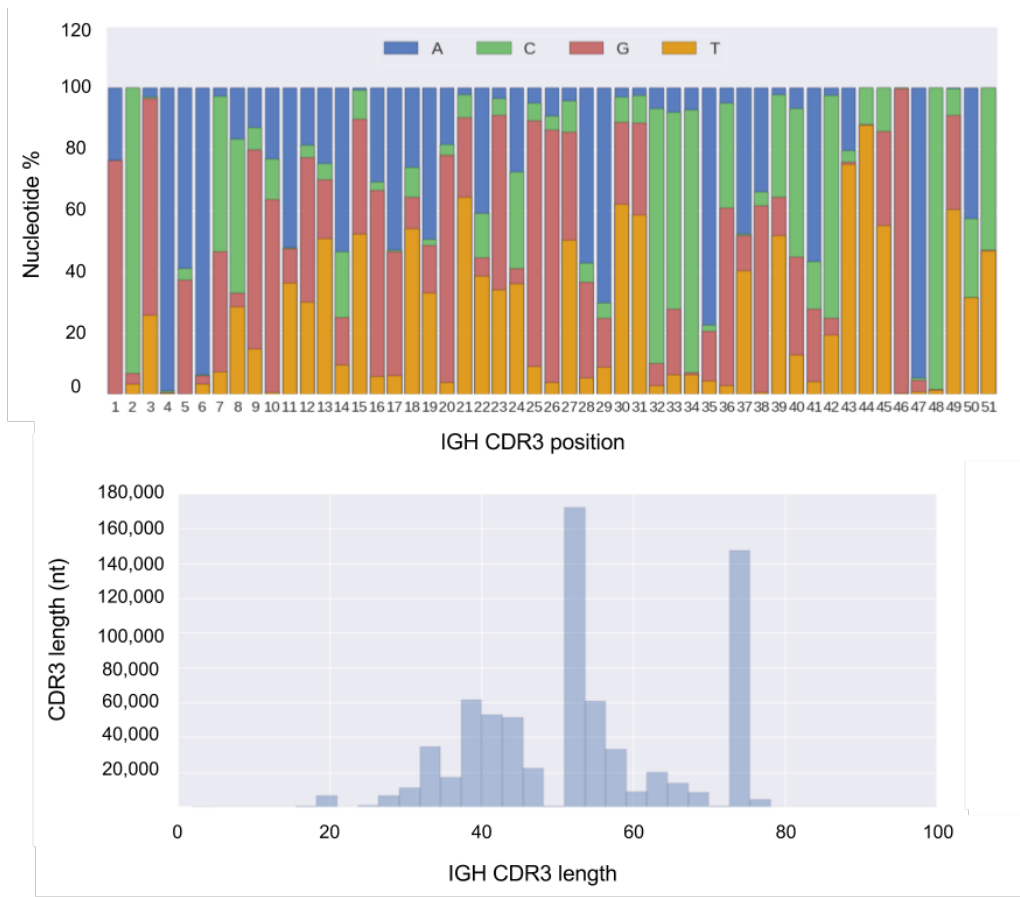

(c)

**Figure 16. Diversity statistics for a repertoire constructed by IGREC for an HIV-infected patient.** (a) Matrix of amino acid substitutions for SHMs. (b) Histogram of distribution of special (SHM to stop codon, insertions, deletions) SHM positions in V segments. (d) Nucleotide variability and length distribution plots of CDRs. IGDIVERSITYANALYZER constructs similar plots for each type of CDR (CDR1/CDR2/CDR3) and each type of chain (IGH/IGK/IGL). It also computes amino acid variability plots.

# Supplemental Material H. Diversity analysis of repertoires constructed by MIXCR and PRESTO

Table 2 presents diversity analysis of repertoires constructed by MIXCR and PRESTO (limited to large clusters with at least 5 reads). Our analysis revealed that both MIXCR and PRESTO accurately reflect important biological features of PBM, AS-, and AS+ datasets.

| MIXCR | | | |
|---|---|---|---|
| | **PBM** | **AS–** | **AS+** |
| *# large clusters* | 5659 | 6630 | 1704 |
| *# reads in large clusters* | 153,343 | 250,599 | 302,568 |
| *# unique CDR3s* | 3610 (64%) | 4634 (70%) | 709 (42%) |
| *# VJ pairs* | 696 (17%) | 527 (13%) | 143 (3%) |
| *max cluster size* | 5271 (3%) | 11,545 (4%) | 89,723 (30%) |
| *max CDR3 abundance* | 15 (0.2%) | 45 (0.7%) | 86 (5%) |
| *SI* | 0.00037 | 0.00050 | 0.0079 |
| *CSI* | 0.00044 | 0.00537 | 0.3054 |
| *CSI / SI* | 1.19 | 10.74 | 38.66 |

| PRESTO | | | |
|---|---|---|---|
| | **PBM** | **AS–** | **AS+** |
| *# large clusters* | 4497 | 6829 | 3570 |
| *# reads in large clusters* | 111,598 | 172,186 | 221,815 |
| *# unique CDR3s* | 2397 (53%) | 4295 (63%) | 893 (25%) |
| *# VJ pairs* | 586 (14%) | 510 (12%) | 135 (3%) |
| *max cluster size* | 2878 (2%) | 6084 (3%) | 40,347 (18%) |

| max CDR3 abundance | 45 (1%) | 188 (2%) | 816 (23%) |
| SI | 0.0007 | 0.0014 | 0.0593 |
| CSI | 0.0009 | 0.0075 | 0.5059 |
| CSI / SI | 1.28 | 5.36 | 8.53 |

**Table 2. Diversity metrics for repertoires constructed by MiXCR (top) and pRESTO (bottom).** The coloring scheme is the same as in Table 1 in the main text.

# Supplemental Material I. Benchmarking parameters

We have benchmarked the repertoire construction tools for non-barcoded datasets (IGREC, PRESTO, and MIXCR) and barcoded datasets (BARCODEDIGREC, PRESTO, and MIGEC). Although various repertoire construction tools use slightly different strategies for read merging, alignment and filtering, we unified the preprocessing step by using IGREC preprocessing (PAIREDREADMERGER and VJ FINDER). After VJ FINDER preprocessing, all input libraries contain Ig-relevant reads that are cropped by the start of the corresponding V gene.

**Benchmarking on non-barcoded datasets.** IGREC has the following parameters: the Hamming graph construction threshold (--TAU), the minimal value of fill-in for dense subgraph finding (--MIN-FILLIN), and the minimal size of super-reads (--MIN-SREAD-SIZE). A *super-read* is defined as a set of identical Rep-seq reads. Super-reads with multiplicity exceeding a threshold represent reliable sequences that should be presented in a final repertoire. Thus, we forbid gluing of such super-reads at the read clustering step.

We launched IGREC with default parameters: --TAU=4, --MIN-SREAD-SIZE=5, and --MIN-FILLIN=0.6 for all benchmarking tests.

Parameters of MIXCR and PRESTO tools are described in Table 3.

**MIXCR parameters**

| name | Parameters | comments |
|---|---|---|
| ALIGN | -f, -g, --noMerge, -p = kaligner2, –species = hsa, -OreadsLayout = Collinear, -OvParameters.geneFeatureToAlign = VTranscript, -OallowPartialAlignments = true | |
| ASSEMBLE | -f, -OassemblingFeatures = FR1Begin:FR4Begin | Since sequences are cropped by the end of CDR3, FR4 region is not present in final |

|  |  | sequences. We selected the specified parameter value since running MiXCR with seemingly more appropriate value FR1Begin:FR4Eend results in a non-stable behavior and often produces an empty repertoire. |
|---|---|---|
| EXPORT CLONES | -f, --no-spaces, -sequence, -count, -readIds |  |

**pRESTO parameters**

| name | parameters | comments |
|---|---|---|
| COLLAPSESEQ | Default parameters | Although this stage can use information about primers, we do not use this information since we want to conduct primer-independent benchmarking. Although this stage can fix unspecified nucleotides ("N"s), but we do not use this feature too, since it is addressed at the preliminary alignment step. |
| SPLITSEQ | Default parameters | The stage uses a threshold parameter (--num=X) that is analogous $minsize_{con}$ in IGREC (discussed in Section 2.2 of the main text). In our experiments, this parameter is not fixed and estimation of its optimal value is a part of benchmarking. |

**Table 3. Benchmarking parameters of MiXCR (top) and pRESTO (bottom) on non-barcoded datasets.**

**Benchmarking on barcoded datasets.** We launched BARCODEDIGREC with default parameters: $\tau_{cluster} = 20$, $\tau_{umi} = 1$, and $\tau_{igrec} = 2$. Since barcoded Rep-seq datasets are often overamplified (giving rise to error-prone super reads), we skipped the procedure of splitting super-reads. We launched IGREC with the default parameters. For benchmarking MIGEC + MIXCR we used the ASSEMBLE stage with the default parameters (for MIGEC) and used MIXCR parameters described in the previous subsection. PRESTO parameters are summarized in Table 4.

| name | parameters |
| --- | --- |
| CLUSTERSETS | Default parameters |
| BUILDCONSENSUS | --prcons 0.6 --maxerror 0.1 --maxgap 0.5 |
| COLLAPSESEQ | --uf PRCONS --cf CONSCOUNT --act sum |

**Table 4. Parameters of PRESTO for benchmarking on barcoded datasets.**

# Supplemental Material J. Benchmarking on SIMULATED COMPLEX dataset

We simulated difficult-to-analyze SIMULATED COMPLEX datasets (modeling highly corrupted clusters) where each input read contains at least one sequencing error. To simulate such libraries, we introduced an additional mismatch at a randomly chosen position in each error-free read in all simple libraries. This procedure increases the average number of errors over all reads from $\lambda$ to $\lambda + e^{-\lambda}$.

In contrast to IGREC, MIXCR and PRESTO failed to construct correct cluster sequences for the SIMULATED COMPLEX datasets. Figure 17 shows that maximal values of *sensitivity+precision* for repertoires constructed by IGREC on the SIMULATED COMPLEX libraries are higher than corresponding values for the SIMULATED SIMPLE libraries.

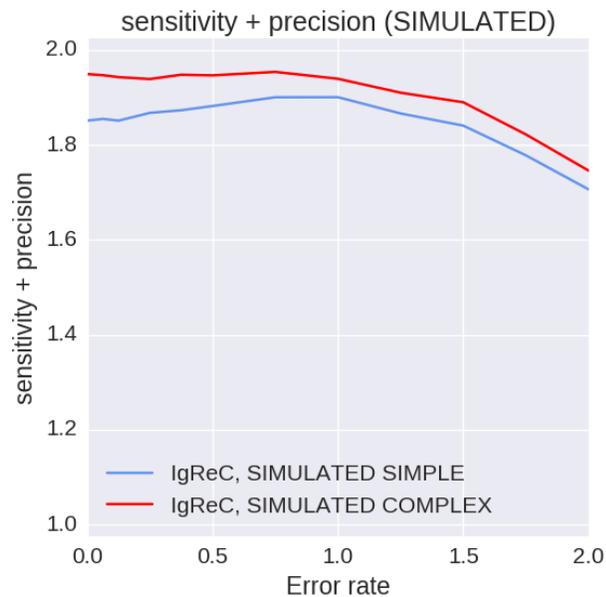

**Figure 17: Sensitivity + precision for various error rate for repertoires constructed** by IGREC for SIMULATED SIMPLE (blue curve) and SIMULATED COMPLEX (red curve) datasets.

# Supplemental Material K. Extending IGSIMULATOR to barcoded Rep-seq datasets

IGSIMULATOR simulates an antibody repertoire and a corresponding Rep-seq library using the ART read simulator (Huang et al. 2012) that models Illumina sequencing errors. However, since ART does not simulate amplification process and other artifacts of Rep-seq datasets, we extended IGSIMULATOR for simulating barcoded Rep-seq libraries.

Our modified version of IGSIMULATOR attaches a randomly and uniformly simulated barcode (with the default length 15 nt) to each receptor sequence in a repertoire simulated by IGSIMULATOR. Afterwards it simulates $N$ amplification cycles (the default value $N = 25$), "amplifying" the set of sequences created at the previous step. Each sequence is amplified with probability $p_{amp}$ (the default value is 0.1). If a sequence is amplified, IGSIMULATOR adds its error-free copy and its error-prone copy to a new set of receptor sequences. IGSIMULATOR generates mismatches introduced into each position of the receptor sequence and barcode with probability $p_{error}$.

Also, IGSIMULATOR simulates chimeric reads by gluing halves of two randomly selected sequences at each amplification cycle. The default rate of chimeras added at each amplification cycle is 0.001. As a result, chimeras account for ≈ 2.5% of the final repertoire.

**Estimating $p_{error}$ using the REAL barcoded Rep-seq dataset.** To simulate a realistic barcoded Rep-seq dataset, we estimated the $p_{error}$ parameter using the REAL dataset. For each barcode, we computed the largest group of identical reads (up to small shifts) and classified it as *dominant* if it (i) contains at least 5 reads and (ii) contains at least 50% more reads than the second largest group. We expect that dominant groups contain error-free reads and thus represent true receptor sequences. For dominant groups, we computed pairwise alignments between the largest group sequence and the remaining reads with the same barcode.

Figure 18 shows the distribution of the Hamming distances computed from these alignments. The average number of mismatches in the computed alignment is close to 2. For simulating Rep-seq libraries, we empirically selected the value $p_{error} = 0.0025$ resulting in 2 errors per read on average.

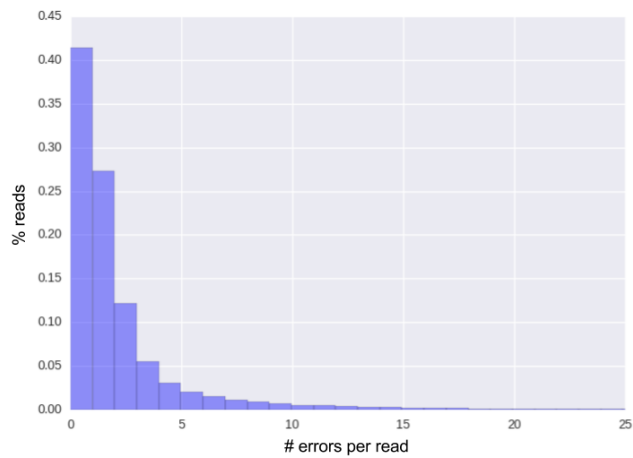

**Figure 18. Distribution of the Hamming distances between sequences of the dominant group in a barcode and error-prone reads.**

# Supplemental Material L. Benchmarking on barcoded Rep-seq datasets with various error rates

We benchmarked BARCODEDIGREC, IGREC (in blind mode), PRESTO, and MIGEC + MIXCR on the SIMULATED BARCODED dataset with various error rate (from 0.5 to 3.75 errors). Figure 19a illustrates that BARCODEDIGREC outperforms other tools on barcoded datasets for the number of errors below 2.6. Surprisingly, IGREC outperforms all other tools in the case of high error rate (more than 2.6 errors). Figure 19b illustrates that all tools are characterized by a dramatic increase in the $minsize_{con}$ threshold when the number of errors exceeds 2. This increase results in a drop of sensitivity (Figure 19c) and a gain of precision (Figure 19d). This observation implies that in the case of datasets with high error rates, it makes sense to analyze rather large clusters, e.g., clusters with at least 10 reads.

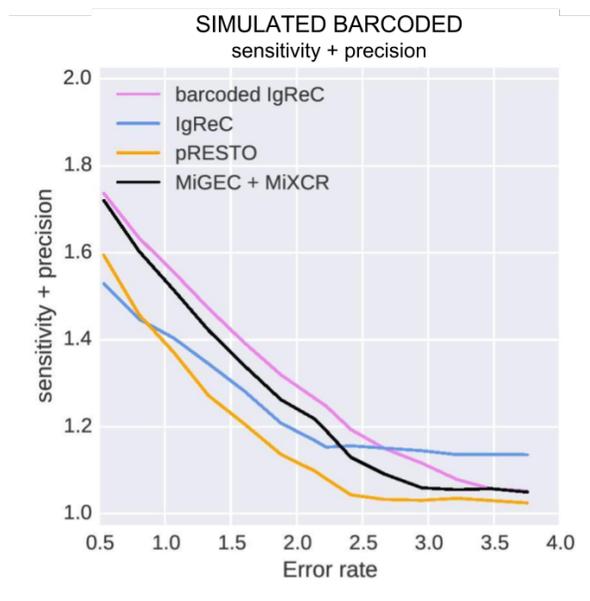

(a)

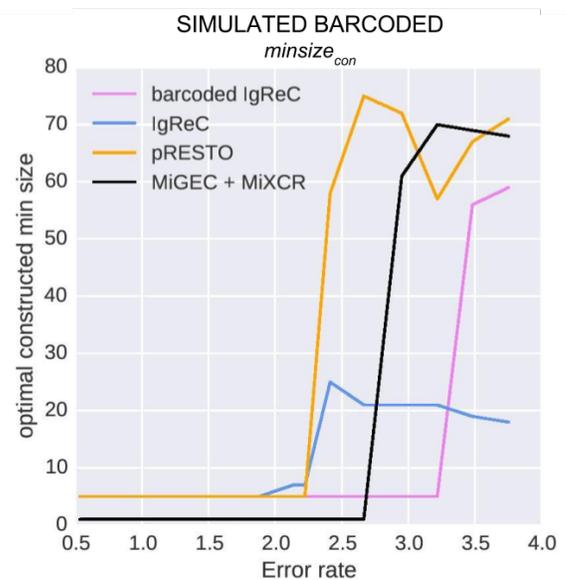

(b)

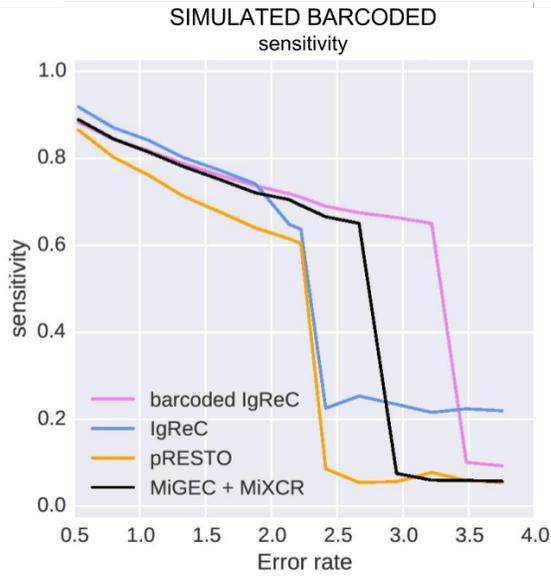 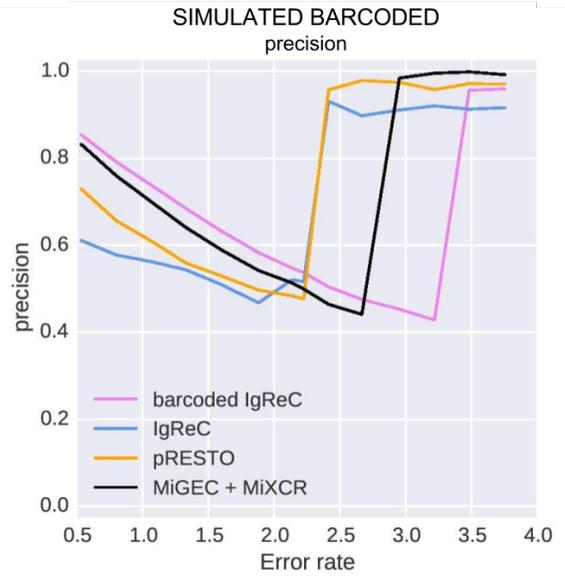

(c) (d)

**Figure 19. Sensitivity and precision depending on the error rates.** For each simulated error rate and each tool, we selected $minsize_{con}$ maximizing sensitivity + precision. (a) Maximal sensitivity + precision versus error rate. (b) $minsize_{con}$ versus error rate. Sensitivity (c) and precision (d) versus error rate for the selected $minsize_{con}$ value.

# Supplemental Material M. Benchmarking IGREC vs IGREPERTOIRECONSTRUCTOR

We benchmarked IGREC (with $\tau = 3$ and $\tau = 4$) against IGREPERTOIRECONSTRUCTOR (with $\tau = 3$). Figure 20 shows the sensitivity-precision plots for the SIMULATED SIMPLE (with 0.5 and 1.0 errors), the SYNTHETIC SIMPLE (with 1.0 errors), and the REAL dataset. For all datasets, except for the SYNTHETIC SIMPLE, IGREC and IGREPERTOIRECONSTRUCTOR demonstrate similar sensitivity and precision. The repertoire constructed by IGREPERTOIRECONSTRUCTOR on the SYNTHETIC dataset has better sensitivity than the repertoires constructed by IGREC. However, despite a small drop in the quality for highly mutated repertoires, IGREC is much faster (several hours as compared to several days for IGREPERTOIRECONSTRUCTOR for a typical Rep-seq dataset), making analysis of large Rep-seq datasets feasible.

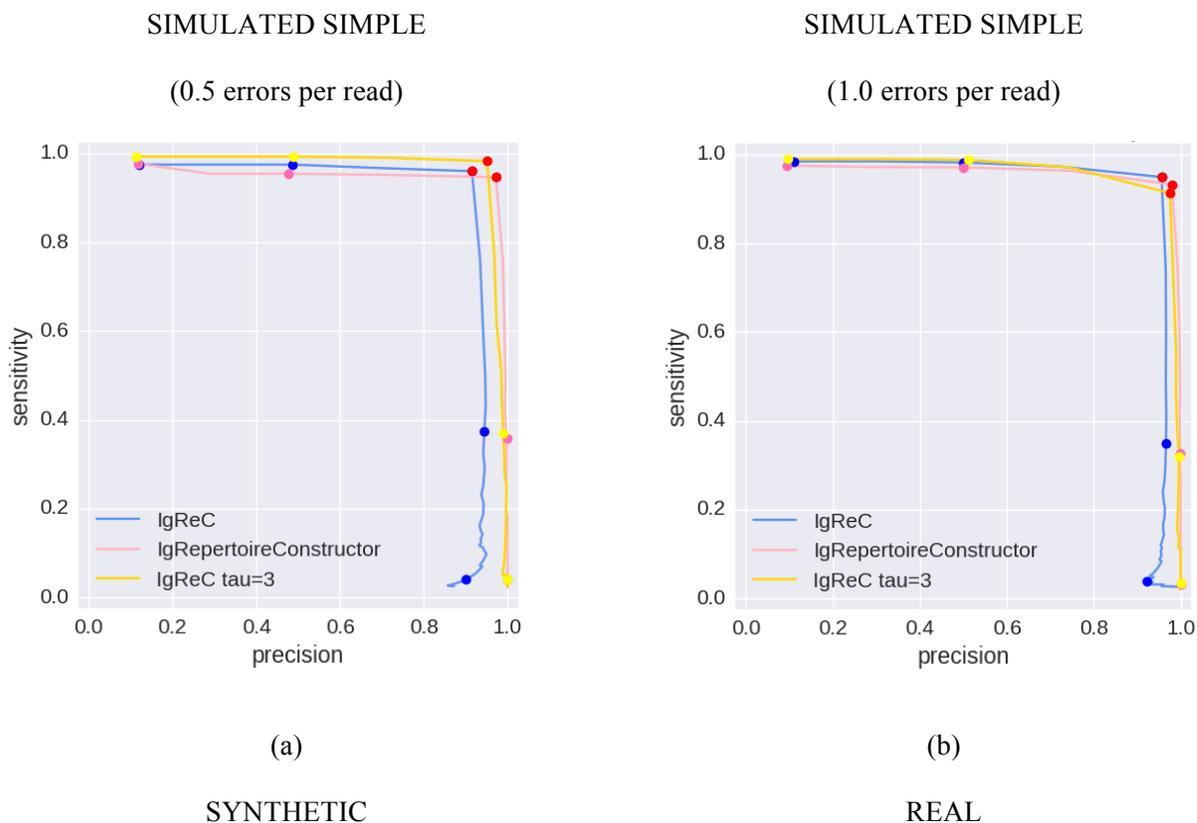

(a) SIMULATED SIMPLE (0.5 errors per read)

(b) SIMULATED SIMPLE (1.0 errors per read)

SYNTHETIC    REAL

(1.0 errors per read)

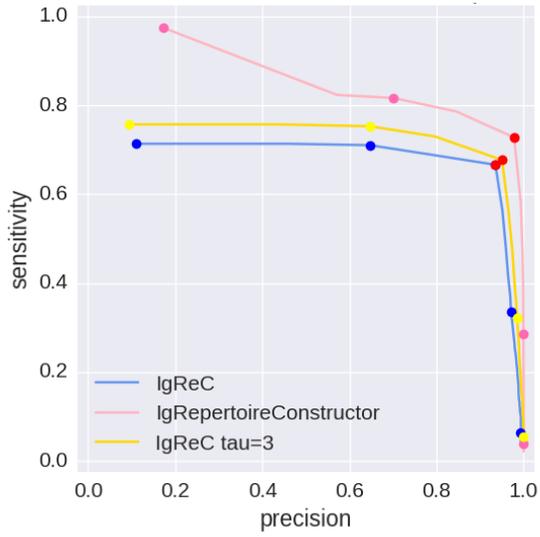 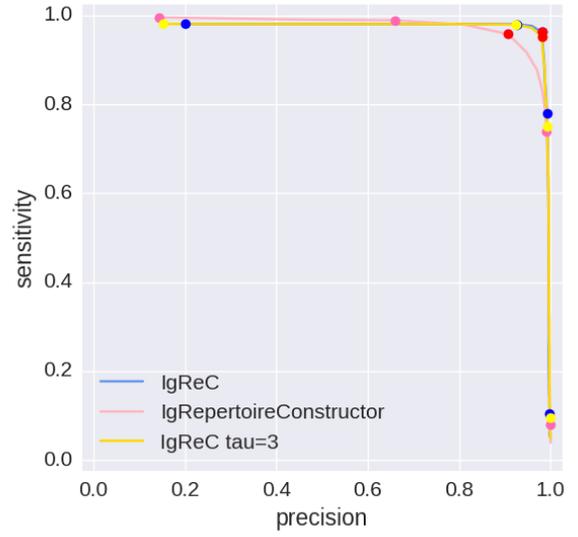

(c) (d)

**Figure 20. Benchmarking of IGREC with default parameters (blue) against IGREC with $\tau = 3$ (yellow) and IGREPERTOIRECONSTRUCTOR (pink) on SIMULATED SIMPLE datasets with 0.5 errors (a) and 1.0 errors (b); SYNTHETIC SIMPLE dataset with 1.0 errors (c); and REAL dataset (d).**

# Supplemental Material N. Correction of barcode errors using BARCODEDIGREC

The quality assessment approach in IGQUAST (based on computing sensitivity and precision) does not reveal the quality of barcoded Rep-seq datasets with respect to their artefacts (chimeric reads, barcode errors, and barcode collisions). To evaluate how BARCODEDIGREC deals with artefacts of barcoded Rep-seq datasets, we analysed barcoded Rep-seq libraries simulated by modified IGSIMULATOR described in Supplemental Material K. Extending IGSIMULATOR to barcoded Rep-seq datasets.

BARCODEDIGREC identified ~30% of chimeric reads (rate of simulated chimeric reads is ~2.5%). The remaining ~70% of simulated chimeric reads form singleton clusters that are discarded after applying the $minsize_{con}$ threshold.

To estimate the accuracy of the barcode collision procedure, we simulated a Rep-seq library with 9 nt-long barcodes. For a repertoire with 43,193 antibody molecules, we generated 39,888 distinct barcodes corresponding to 3305 barcode collisions. BARCODEDIGREC resolved 96.2% of them (125 collisions remained unresolved). Simulation with more realistic collision rate resulted in 100% resolution of barcode collisions.

We also analysed a simulated barcoded library with erroneous barcodes containing 480,078 reads and 43,193 distinct barcodes (47,972 of the reads or approximately for ~10% had erroneous barcodes). BARCODEDIGREC identified and corrected 68% of reads with erroneous barcodes (i.e., glued them to a cluster with correct barcode). The remaining reads with erroneous barcodes (32%) form singleton clusters.